\documentclass[prb,showpacs,twocolumn,superscriptaddress]{revtex4}
\usepackage{amsmath,graphicx,latexsym,natbib,bm,psfrag}

\def\msm#1{\marginpar{\small MS: #1}}
\def\jjm#1{\marginpar{\small JJ: #1}}
\def\pa#1{\marginpar{\small PA: #1}}

\def\msm#1{}
\def\jjm#1{}
\def\pa#1{}

\begin{document}

\title{Band alignment at metal/ferroelectric interfaces: 
insights and artifacts from first principles}

\author{Massimiliano Stengel}
\affiliation{Institut de Ci\`encia de Materials de Barcelona (ICMAB-CSIC), 
              Campus UAB, 08193 Bellaterra, Spain}
\author{ Pablo Aguado-Puente}
\affiliation{ Departamento de Ciencias de la Tierra y
              F\'{\i}sica de la Materia Condensada, Universidad de Cantabria,
              Avda. de los Castros s/n, 39005 Santander, Spain}
\author{Nicola A. Spaldin}
\affiliation{Department of Materials, ETH Zurich, Wolfgang-Pauli-Strasse 27, 
8093 Zurich, Switzerland }
\author{ Javier Junquera }
\affiliation{ Departamento de Ciencias de la Tierra y
              F\'{\i}sica de la Materia Condensada, Universidad de Cantabria,
              Avda. de los Castros s/n, 39005 Santander, Spain}

\date{\today}

\begin{abstract}
 Based on recent advances in first-principles theory, we develop a general model of 
 the band offset at metal/ferroelectric interfaces. 
 We show that, depending on the polarization of the film, a pathological regime 
 might occur where the metallic carriers populate the energy bands of the
 insulator, making it metallic.
 As the most common approximations of density functional
 theory are affected by a systematic underestimation of the fundamental band 
 gap of insulators, this scenario is likely to be an artifact of the simulation.
 We provide a number of rigorous criteria, together with extensive practical examples, 
 to systematically identify this problematic situation in the calculated electronic
 and structural properties of ferroelectric systems.
 We discuss our findings in the context of earlier literature studies, 
 where the issues described in this work have often been overlooked.
 We also discuss formal analogies to the physics of polarity compensation at
 LaAlO$_3$/SrTiO$_3$ interfaces, and suggest promising avenues for future 
 research.
\end{abstract}


\maketitle

\section{Introduction}
\label{sec:intro}

 Advances in oxide thin film growth techniques over the last ten years have led
 to the fabrication of many novel oxide-based metal-insulator heterostructures
 with a dizzying range of functionalities.
 Not only are the current technological limits of information 
 storage density and
 speed being pushed forward by the use of, e.g., nanoscale ferroelectric
 memories,~\cite{Scott-89, Auciello-98, Scott-book, Dawber:2005, 
 Scott-06, Scott-07} but entirely
 new concepts in device applications are
 also emerging, in which the electrical and the magnetic degrees of freedom are
 both present within the same active element and strongly 
 coupled.~\cite{Fiebig-05,Eerenstein-06}
 Examples of this trend include thin-film capacitors,~\cite{Dawber:2005}
 strongly correlated
 field-effect devices,~\cite{Ahn-03}
 and magnetic/ferroelectric tunnel 
 junctions.\cite{Tsymbal/Kohlstedt:2006, Maksymovych-09, Garcia-09, Garcia-10}

 Density functional theory (DFT) methods, either within the
 local density (LDA) or generalized gradient (GGA) approximation,
 have been an invaluable tool in achieving a fundamental understanding of
 this class of systems,\cite{Dawber:2005,Ghosez:2006,Junquera_review:2008}
 particularly with recent developments which allow the application of finite
 electric fields to periodic solids or layered heterostructures.
 \cite{Souza/Iniguez/Vanderbilt:2002,Umari-02,Souza-04,Stengel/Spaldin:2007,
 Stengel-09.2}
 However, since this domain of research is relatively new, it is
 important to identify, in addition to the virtues, also the
 limitations of DFT that are specific to metal/ferroelectric interfaces,
 and that when overlooked might lead to erroneous physical conclusions.

 For most practical applications, a capacitor must be insulating to DC current;
 transmission of electrons via non-zero conductivity and/or direct tunneling
 (leakage) is generally an undesirable source of heating and power consumption.
 At the quantum mechanical level, the insulating properties of a capacitor are
 guaranteed by the presence of a dielectric film with a finite band gap at
 the Fermi level, where propagation of the metallic conduction electrons
 is forbidden.
 In the language of semiconductor physics, we can alternatively say that
 both Schottky barrier heights (SBH), respectively $\phi_n$ and $\phi_p$ 
 for electrons and holes, need to be positive for the device to behave 
 as a capacitor. 
 (By convention we assume that, if the Fermi level of the metal lies in 
 the gap of the insulator, both $\phi_n$ and $\phi_p$ are positive.)

 If, on the contrary, either $\phi_p$ or $\phi_n$ is negative,
 injection of holes or electrons into the dielectric becomes 
 energetically favorable and the device behaves instead as an Ohmic contact.
 Most importantly, at such a junction there is necessarily
 (at thermodynamic equilibrium) a spill-out of charge from the metal to
 the insulator, as the system re-equilibrates
 the chemical potential of the free carriers on either side.
 Such intrinsic space charge induces metallicity (by intrinsic doping)
 in the dielectric film, and overall profoundly alters the
 electronic and structural properties of the interface.

 While in principle the charge spillage might be a real physical feature
 of a given system, there are several arguments that advise caution in
 the interpretation of DFT calculations where this effect is found.
 The use of an approximate functional to model the exchange and
 correlation energy, such as LDA or GGA, generally produces 
 severe and systematic errors in the values of  $\phi_p$ and $\phi_n$,
 which can be generally traced back to the well-known band-gap 
 problem.~\cite{Perdew-83,Sham-83}
 This implies that finding a negative value of either $\phi_p$ or $\phi_n$ 
 is unlikely to be a robust result of an LDA or GGA calculation.
 Furthermore, the total amount of spilled-out charge depends on the 
 DFT values of $\phi_p$ and $\phi_n$ (the more negative the SBH, the larger 
 the number of states of the insulator that cross the Fermi level).
 This means that, in such a pathological regime,
 the error in $\phi_p$ or $\phi_n$ will directly propagate to the 
 charge density, and potentially affect a number of fundamental 
 ground-state properties of the interface. 
 In order to avoid undesirable artifacts in the DFT results, it is therefore 
 crucial to clearly identify whether this scenario applies
 to a given interface calculation.

 Such an analysis is not entirely straightforward, as physics governing 
 the band alignment in a ferroelectric capacitor significantly departs
 from the well-established concepts of semiconductor physics.
 First, the imperfect screening at the electrode interface produces 
 a potential drop~\cite{nature_2006,Junquera_review:2008} that is roughly 
 linear in the polarization $P$,\cite{Junquera/Ghosez:2003} and modifies the 
 lineup between the bands of the insulator and the Fermi level 
 of the metal.~\cite{Stengel-09.2}
 This phenomenon, central to the physics of ferroelectric
 capacitors, has important implications for the stability
 of a monodomain polar state,\cite{Junquera_review:2008} and for
 devices based on the tunneling electroresistance 
 effect.~\cite{Zhuravlev_et_al:2005}
 Second, the residual ``depolarizing'' electric field 
 produces a linearly increasing electrostatic potential in the film.
 This prevents a precise determination of the band 
 lineup,~\cite{Stengel-09.2} 
 as a proper (and physically meaningful) definition of the latter requires a
 macroscopically constant reference energy in the insulating region.
 Third, the marked covalent character of bonding in perovskites produces 
 non-trivial changes in the band structure of the insulator, depending 
 on the magnitude of the polar distortion.
 This further complicates the extraction of an accurate band lineup by
 means of standard first-principles procedures, as the bulk reference calculation
 needs to accurately match the \emph{electrical}, in addition to the
 mechanical, boundary conditions of the film.
 Finally, and most importantly, one must keep in mind that all these new physical 
 ingredients may coexist with the more traditional features that are typical of
 metal/semiconductor interfaces, e.g. the phenomenon of metal-induced gap states
 (MIGS).\cite{Heine-65}
 To guide future works in this field, and to build a firm 
 theoretical basis for the
 interpretation of the experiments, it is becoming increasingly 
 urgent to rationalize 
 all these many competing effects into a coherent picture, where the limitations
 of the current simulation methods can be clearly drawn.

 Here we develop a general and intuitive model of 
 the band offset at a ferroelectric/metal interface, and its dependence on the
 polarization.
 We identify two qualitatively distinct regimes, corresponding to (i) that
 of a normal Schottky alignment and (ii) that of a pathological Ohmic junction.
 We demonstrate the artifacts typically associated with (ii) by 
 performing extensive calculations of technologically relevant 
 ferroelectric/metal interfaces.
 We discuss the relevant literature works, pointing out
 those where our results suggest a revision of the currently 
 accepted interpretation.
 We further identify a direct relationship between the pathological
 Ohmic regime and the physics of ``electronic 
 reconstruction''~\cite{Okamoto-04} at polar oxide interfaces
 such as LaAlO$_3$/SrTiO$_3$, and trace a viable route towards a 
 unified description of these two phenomena.
 Finally, we discuss a number of viable methodological perspectives to
 overcome the limitations of DFT illustrated in this work.

 The paper is organized as follows: 
 In Sec.~\ref{sec:theory} we develop our theoretical model of the
 band offset at a ferroelectric/metal interface, illustrating the 
 main consequences of a ``pathological'' band alignment.
 In Sec.~\ref{sec:methods} we present a self-contained
 overview of the theoretical methods we use to detect such 
 features in a first-principles calculation. 
 In Sec.~\ref{sec:results-para} we 
 present the results of our simulations for paraelectric capacitors, 
 by comparing non-pathological (PbTiO$_3$/SrRuO$_3$ and BaTiO$_3$/SrRuO$_3$) 
 and pathological cases (KNbO$_3$/SrRuO$_3$ and BaTiO$_3$/Pt).
 In Sec.~\ref{sec:results-ferro} we demonstrate that the two cases which we 
 find to be non-pathological in the paraelectric configuration indeed become 
 pathological when the polarized ferroelectric state is fully relaxed.
 In Sec.~\ref{sec:discussion} we discuss the implications of this work
 with respect to the existing literature on the subject.
%
 Finally, in Sec.~\ref{sec:conclusions} we present 
 our conclusions and outlook for future research.

\section{General theory of the band offset}
\label{sec:theory}

\subsection{Metal/semiconductor interfaces}
\label{sec:metsemi}

The Schottky barrier, a rectifying barrier for electrical 
conduction across a metal/semiconductor junction, is of vital
importance for the operation of any modern electronic device.
For the case of an $n$-type semiconductor, the Schottky barrier 
height is the energy difference between the conduction band 
minimum and the Fermi level across the interface, and we indicate
it as $\phi_n$.
The nature of the microscopic mechanisms governing the magnitude
of $\phi_n$ has troubled scientists for several decades.
In spite of the ongoing debates, it seems to be widely accepted now 
that, while bulk material properties certainly play a substantial 
role, $\phi_n$ is best understood as a genuine \emph{interface} 
property.
This is in agreement with the intuitive picture one gets from quantum 
mechanics: the charge rearrangement due to chemical bonding at the
interface produces an interface dipole, and this will uniquely 
determine the offset between the energy bands of the insulator and
the Fermi level of the metal.

 \begin{figure}
    \begin{center}
    \includegraphics[width=3.0in] {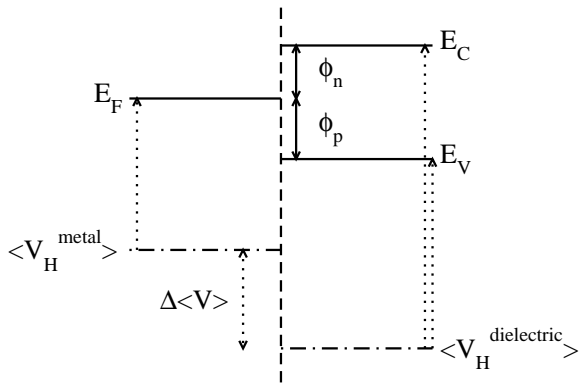}
       \caption{ 
         Schematic representation of the band offset at a metal/insulator
         junction, illustrating the main quantities discussed in the text.
         }
       \label{fig:offset}
    \end{center}
 \end{figure}

To be more specific, it is useful to consider the electrostatic 
Hartree potential at the interface between two semi-infinite solids,
\begin{equation}
V_{\rm H} ({\bf r}) = \int \frac{\rho ({\bf r'})}{|{\bf r}-{\bf r'}|} d^3 r',
\end{equation}
where $\rho(\bf{r})$ is the total charge density (including electrons
and nuclei).
$V_{\rm H}$ is a rapidly varying function of the position, reflecting the
underlying atomic structure.
In order to filter out the large oscillations and preserve only 
those features that are relevant on a macroscopic scale, it is
convenient to apply an averaging procedure.~\cite{Baldereschi-88,Colombo-91}
This consists of (i) performing a global average of $V_{\rm H}({\bf r})$ over 
planes parallel to the interface, and (ii) convoluting the resulting
one-dimensional function with a Fourier filter to suppress the 
high spatial frequency components. (See Ref.~\onlinecite{Junquera-07}
for a detailed description of the method, and Ref.~\onlinecite{Franciosi-96}
for an extensive review of its applications to SBH calculations.)
After this ``nanosmoothing''~\cite{Junquera-07} procedure, the
doubly-averaged $\overline{ \overline{V}}_{\rm H}(z)$ reduces to a step function,
from which we can extract the electrostatic 
\emph{lineup term},~\cite{Baldereschi-88,Colombo-91}
\begin{equation}
\Delta \langle V \rangle = \langle V_{\rm H}^{\rm dielectric} \rangle - 
\langle V_{\rm H}^{\rm metal} \rangle,
\label{eq:deltav}
\end{equation}
which includes all the physics of the interface dipole formation. 
[$\langle V_{\rm H}^{\rm dielectric} \rangle$ and $\langle V_{\rm H}^{\rm metal} \rangle$
are the asymptotic values of $\overline{ \overline{V}}_{\rm H}(z)$ far from the interface.]
To determine the band offsets from $\Delta \langle V \rangle$ it is then 
necessary to know how the bulk energy bands of the insulator and the Fermi 
level of the metal are related to their respective average electrostatic
potential. 
%
In full generality, one can write
 \begin{subequations}
    \begin{align}
       \phi_{p} & = E_{\rm V} - E_{\rm F} + \Delta \langle V \rangle ,
       \label{eq:phip}
       \\
       \phi_{n} & = E_{\rm C} - E_{\rm F} + \Delta \langle V \rangle .
       \label{eq:phin}
    \end{align}
 \end{subequations}
 $E_{\rm V}$, $E_{\rm C}$ and $E_{\rm F}$ are usually referred to as 
 {\it the band structure term},~\cite{Baldereschi-88,Colombo-91} and
 are bulk properties of the two materials. They are defined as 
 the energy positions of the valence ($E_{\rm V}$) and conduction ($E_{\rm C}$)
 band edges of the insulator, and the Fermi level of the metal ($E_{\rm F}$),
 all referred to the average $\langle V_{\rm H} \rangle$ in the respective bulk 
 (see Fig.~\ref{fig:offset}).

In Sec.~\ref{sec:methods} we provide further details of the standard computational
procedures used to calculate these quantities in practice. In the following Section
we discuss how the above theory needs to be revised and extended in the case of
metal/ferroelectric interfaces.

\subsection{Metal/ferroelectric interfaces}
\label{sec:metferro}

Ferroelectric materials entail a new degree of freedom, the macroscopic
polarization $P$, which is absent in the semiconductor case. It is natural
then to expect that the above picture of the band offset at metal/insulator 
interfaces may need to be extended to take this new variable into account. 
In the following, we discuss how $P$ affects both the lineup and the 
band-structure terms in Eqs.~(\ref{eq:phip}) and~(\ref{eq:phin}).

\subsubsection{Lineup term}

We represent a simple ferroelectric material as a non-linear dielectric, 
which in bulk is characterized by an internal energy $U_{\rm b}$ per
unit cell of the form
\begin{equation}
U_{\rm b}(D) = A_0 + A_2 D^2 + A_4 D^4 + \mathcal{O}(D^6).
\label{eq:uofd}
\end{equation}
Here $D$ is the electric displacement field, $A_0$ is an arbitrary reference
energy, $A_2$ is \emph{negative} and the highest expansion coefficient  
positive. (As we are concerned with the essentially one-dimensional case of 
a parallel-plate capacitor, 
we only consider the component of the ${\bf D}$ vector
that is normal to the interface plane, indicated as $D$ henceforth.) 
The $A_{0,2,4,\ldots}$ coefficients 
implicitly contain all the complexity of the microscopic physics, and can be
calculated from first principles using the methods of Ref.~\onlinecite{fixedd}.
It follows from elementary electrostatics~\cite{fixedd} that the internal 
electric field, $\mathcal{E}(D)$, is the derivative of $U(D)$,
\begin{equation}
\mathcal{E}_{\rm b}(D) = \frac{1}{\Omega} \frac{dU_{\rm b}(D)}{dD},
\label{eq:eofd}
\end{equation}
where $\Omega$ is the cell volume.

 \begin{figure}
    \begin{center}
    \includegraphics[width=2.5in] {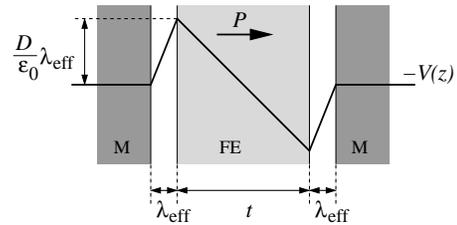}
       \caption{ 
         Schematic representation of a symmetric short-circuited ferroelectric 
         capacitor in a polarized configuration
         within the imperfect-screening model. $t$ is the
         thickness of the ferroelectric film. 
         M and FE represent, respectively, the metal electrode
         and the ferroelectric film. Both materials are assumed to be 
         separated by a vacuum layer of thickness $\lambda_{\rm eff}$.
         The thick solid line indicates the opposite of the electrostatic
         potential, $-V(z)$.}
       \label{fig:depolar}
    \end{center}
 \end{figure}

The electrostatics of a parallel-plate capacitor configuration can be well
described~\cite{nature_2006,nature_mat,Junquera/Ghosez:2003} within the
\emph{imperfect screening} model, as sketched in Fig.~\ref{fig:depolar}.
The $N$-layer thick ferroelectric film can be thought of as separated
from the ideal metal electrode by a thin layer of vacuum, of thickness
$\lambda_{\rm eff}$. $\lambda_{\rm eff}$ is an ``effective
screening length'' that takes into account all the microscopic details of
the interface dipole response to the polarization,~\cite{Junquera/Ghosez:2003}
including electronic and chemical bonding effects.~\cite{nature_mat}
 At the interface between the ferroelectric and the vacuum layer
 $D$ must be preserved.
 Therefore, an homogeneous electric field appears inside the vacuum layer,
 of magnitude $\mathcal{E}_{\rm vac} = D/\epsilon_{0}$.
 Recalling that the energy density of a static electric field $\mathcal{E}$
 in vacuum is $u=\epsilon_0 \mathcal{E}^2/2= D^2/2\epsilon_0$, the energy 
 of the $N$-layer thick ferroelectric film can 
 then be written as 
\begin{equation}
U_N(D) = N U_{\rm b}(D) + 2S \lambda_{\rm eff} \frac{D^2}{2\epsilon_0},
\label{eq:ucapa}
\end{equation}
where $S$ is the surface cell area. [Note that two symmetric electrodes of
equal $\lambda_{\rm eff}$ are considered in Eq.~(\ref{eq:ucapa}).]
The second important consequence of a non-zero $\lambda_{\rm eff}$ is that the
lineup term, Eq.~(\ref{eq:deltav}), now linearly depends on the external 
parameter $D$, due to the additional potential drop at the interface,
that can be computed as the product of the electric field within the 
vacuum layer times its width,
\begin{equation}
\Delta \langle V \rangle(D) = \Delta \langle V \rangle(0) + 
\lambda_{\rm eff} \frac{D}{\epsilon_0}.
\label{eq:deltavd}
\end{equation}
[It is worth noting that, whenever $\mathcal{E}_{\rm b}(D) \neq 0$, at the microscopic
level $\Delta \langle V \rangle(D)$ contains an intrinsic arbitrariness; furthermore,
in such a case it is no longer justified to think in terms of an ``isolated'' interface
between two semi-infinite solids.
Techniques to deal with these issues in practical calculations
are described in Ref.~\onlinecite{Stengel-09.2}.]

 \begin{table}
    \begin{ruledtabular}
       \begin{tabular}{crrr}
                                                       &
          \multicolumn{1}{c}{$D_{\rm S}$ (C/m$^2$)} &
          \multicolumn{1}{c}{$\lambda_{\rm eff}$ (\AA)}    &
          \multicolumn{1}{c}{$\Delta \phi$ (V)}        \\
          \hline
          BaTiO$_3$ &  0.39 & 0.20 & 1.8 \\
          PbTiO$_3$ &  0.75 & 0.15 & 2.6 \\
       \end{tabular}
       \caption{ Estimation of the change
                 in the lineup term $\Delta \phi$
                 of typical ferroelectric capacitors upon polarization reversal. 
                 $D_{\rm S}$ is the
                 bulk spontaneous polarization of the ferroelectric material.
                 $\lambda_{\rm eff}$ were calculated in 
                 Ref. \onlinecite{nature_mat} for capacitors with
                 SrRuO$_3$ electrodes.
                }
       \label{tab:deltap}
    \end{ruledtabular}
 \end{table}

To give a more quantitative flavor of the impact of this $D$-dependence in real
systems, we can use the values of $\lambda_{\rm eff}$ reported in the literature 
for PbTiO$_3$/SrRuO$_3$ and BaTiO$_3$/SrRuO$_3$ capacitors.
Upon polarization reversal, the interface lineup term $\Delta \langle V \rangle$
will undergo a variation corresponding to 
\begin{equation}
\Delta \phi =  
\Delta \langle V \rangle(D_{\rm S}) - \Delta \langle V \rangle(-D_{\rm S}) = 
2\lambda_{\rm eff} \frac{D_{\rm S}}{\epsilon_0},
\end{equation}
where $D_{\rm S}$ is the spontaneous polarization of the ferroelectric material
(in the spontaneous configuration the internal electric field within the
ferroelectric, $\mathcal{E}_{\rm b}$, vanishes and $D_{\rm S}$ equals the 
spontaneous polarization.)
The values reported in Table~\ref{tab:deltap} indicate that this effect 
can be rather large, of the order of 1-2 eV, even for ideal defect-free interfaces.

\subsubsection{Band-structure term}
\label{sec:bandstr}

The polar displacements in the ferroelectric film modify not only the lineup
term, but also the bulk band-structure term. This is most easily understood 
by recalling the role played by covalent bonding in the ferroelectric instability
of perovskite titanates. Hybridization effects between the cation $3d$ states 
and the oxygen $2p$ states are intimately linked to the off-centering of the
Ti sublattice. This implies that the polar distortions can significantly
modify both the conduction and valence band structure. For example, in both
BaTiO$_3$ and PbTiO$_3$ the fundamentamental gap increases when going 
from the centrosymmetric cubic structure to the polar tetragonal phase.
Using the arguments of Ref.~\onlinecite{Stengel-09.2}, we can think of
a continuous dependence of both $E_{\rm V}$ and $E_{\rm C}$, respectively
in Eq.~(\ref{eq:phip}) and Eq.~(\ref{eq:phin}), on the electric
displacement $D$. The Fermi level $E_{\rm F}$, of course, remains fixed as the
electric displacement does not affect the bulk of the metallic electrode.
In summary, the general expression for the $n$-type SBH at a metal/ferroelectric
interface (an analogous expression follows for the $p$-type one) is
\begin{equation}
  \phi_{n}(D) = E_{\rm C}(D) - E_{\rm F} + \Delta \langle V \rangle(D),
 \label{eq:phid}
\end{equation}
where at the lowest order $E_{\rm C}$ is quadratic in $D$ (the linear order
is forbidden by symmetry), and in most cases of interest 
$\Delta \langle V \rangle(D)$ can be approximated by a linear 
function as in Eq.~(\ref{eq:deltavd}).
In the following, we shall elaborate on this expression and identify 
a new, qualitatively different regime, with important implications
for the physics of the interface.

\subsection{Ferroelectric capacitors in a pathological regime}

Equation~(\ref{eq:phid}) implies that $\phi_{n}(D)$ might become
\emph{negative} for some values of $D$. From the point of
view of first-principles calculations, already by looking at the values
of Table~\ref{tab:deltap} we can be reasonably sure that this \emph{will}
happen at the PbTiO$_3$/SrRuO$_3$ interface: 2.6 eV is already larger 
than the LDA gap of PbTiO$_3$ in the ferroelectric phase ($\sim$2.0 eV).
This possibility has been almost
systematically overlooked in the literature. As this is a central point of
this work, we shall illustrate in detail the consequences of such a regime,
and explain why we regard it as ``pathological''.
We discuss in the following two possible occurrences of this scenario:
(i) $\phi_{n}$ is negative already in the paraelectric configuration 
at $D=0$ and (ii) $\phi_{n}$ is positive at $D=0$ but becomes negative
at some value of $|D|<D_{\rm S}$.

\subsubsection{The centrosymmetric case}
\label{sec:paraspill}

\begin{figure}
    \begin{center}
    \includegraphics[width=2.8in] {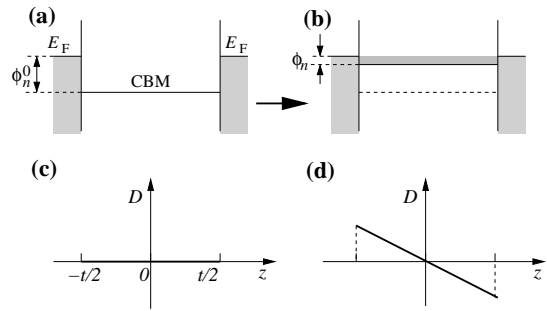}
       \caption{ 
         Schematic representation of the effect of free-charge
         redistribution onto the band diagram of a paraelectric capacitor
         with a negative $\phi_{n}$.
         (a) band alignment under perfect \emph{interface} screening
         (i. e. when $\rho_{\rm free}$ vanishes), and
         (b) after charge spill-out and electrostatic re-equilibration.
         The corresponding profiles of the electric displacement field
         within the ferroelectric films are displayed in panels (c) and (d),
         obtained after integrating Eq. (\ref{eq:maxwell}).
         }
       \label{fig:prespill}
    \end{center}
 \end{figure}

We start with a capacitor in the reference paraelectric structure with
two symmetric electrodes, and we hypothesize that, for whatever physical 
reason, the interface dipole that forms between the metal and the film 
leads to a \emph{negative} $\phi_n$. (Similar arguments apply to the 
case, not explicitly discussed here, of a negative $\phi_p$.)
As the quantum states of the conduction band of the film lie at lower 
energy than the Fermi level of the metal, the former will be filled 
up to $E_{\rm F}$, leading to a nonzero free charge density, 
$\rho_{\rm free}$, in the film. 
Neglecting quantum confinement effects, we can use the Thomas-Fermi model
and treat the free charge distribution as macroscopically uniform. 
Within this approximation, $\rho_{\rm free}$ is exactly given in terms of 
$\phi_n$ and the electronic density of states of the bulk insulator, $\rho_{\rm b}(E)$,
in a vicinity of the conduction band bottom, $E_{\rm C}$,
\begin{equation}
\rho_{\rm free} = -\frac{e}{\Omega} \int_{E_{\rm C}}^{E_{\rm C}-e\phi_n} \rho_{\rm b}(E) dE.
\label{eq:rhof}
\end{equation}
This additional charge density, superimposed on an otherwise charge-neutral
insulating film, will produce a strong electrostatic perturbation in the
system.
For example, if such a charge rearrangement occurred in vacuum, the Poisson equation
 \begin{equation}
    \frac{d^{2}V(z)}{dz^{2}} = -\frac{\rho_{\rm free}}{\epsilon_{0}},
 \end{equation}
would imply a parabolic potential of the form
 \begin{equation}
    V(z) = -\frac{\rho_{\rm free}}{2 \epsilon_{0}} z^{2}.
    \label{eq:pertpot}
 \end{equation}
(We assume that $z=0$ corresponds to the center of the ferroelectric
film.)
Throughout this work, we shall assume that the interface is oriented
along the $z$ axis, and each material is periodic in the plane
parallel to the interface, referred to as the $(x,y)$ plane.
As typical ferroelectric materials are exceptionally good dielectrics, 
in a first approximation we can assume that $V(z)$ will be perfectly
screened by the polar displacements of the lattice.
However, this does not mean that electrostatics has no consequences -- 
quite the contrary. Macroscopic Maxwell equations in materials indeed 
dictate that
\begin{equation}
\frac{dD(z)}{dz} = \rho_{\rm free}.
\label{eq:maxwell}
\end{equation}
Hence, if we assume perfect \emph{bulk} screening, we have 
$\mathcal{E}(z)=0$, $D(z)=\epsilon_0 \mathcal{E}(z) + P(z)= P(z)$ 
and, after integrating Eq.~(\ref{eq:maxwell}), $P(z)=\rho_{\rm free} z$. 
So, since the sign of the electronic charge and $\rho_{\rm free}$ 
are negative within our convention, 
we have a non-uniform and linearly \emph{decreasing} polarization
in the ferroelectric film [see Fig.~\ref{fig:prespill}(d)].
This means that, at the film boundaries ($z=\pm t/2$, where $t$ is the 
thickness), the local electric displacement has now opposite values, 
proportional to the \emph{total} amount of free charge that was transferred,
\begin{equation}
D(-\frac{t}{2}) =  -\frac{t}{2} \rho_{\rm free}, \qquad D(\frac{t}{2}) =  \frac{t}{2} \rho_{\rm free}.
\end{equation}
Of course, the band offset at the interface depends on the \emph{local}
value of $D$ in the film region adjacent to the interface, so
$\phi_n$ will be consequently shifted in energy according 
to Eq.~(\ref{eq:phid}). We can expect that for small $D$ 
values the (quadratic) polarization effects on the band 
structure will be less important than the (linear) dependence 
of the lineup term on $D$. (Note that the presence of additional 
charge in the conduction band might also alter the bandstructure
term, e.g. through on-site Coulomb repulsions or other exchange 
and correlation effects; in the limit of weak correlations we 
expect these to be even smaller and essentially irrelevant for 
this discussion.)
Therefore, we approximate
Eq.~(\ref{eq:phid}) with Eq.~(\ref{eq:deltavd}), and write 
\begin{equation}
\phi_n = \phi_n^0 - \frac{\lambda_{\rm eff} D}{\epsilon_0} = \phi_n^0 - 
 \frac{t \lambda_{\rm eff} \rho_{\rm free}}{2\epsilon_0}.
\label{eq:phipr}
\end{equation}
[The minus sign comes from the fact that at the $z<0$ interface,
which is the one for which Eq.~(\ref{eq:deltavd}) is valid within
our conventions, $D$ is negative.]
In turn, the new $\phi_n$ will modify $\rho_{\rm free}$ through 
Eq.~(\ref{eq:rhof}). For some value of $\phi_n$, Eq.~(\ref{eq:rhof}) and
Eq.~(\ref{eq:phipr}) will be mutually self-consistent and the system will reach 
electrostatic equilibrium. 
This can be expressed through an integral equation where we have eliminated
$\rho_{\rm free}$,
\begin{equation}
\frac{e}{\Omega}  \int_{E_{\rm C}}^{E_{\rm C}-e\phi_n}  \rho_{\rm b}(E) dE = 
\frac{2\epsilon_0(\phi_n-\phi_n^0) }{t \lambda_{\rm eff}}.
\end{equation}
%

To qualitatively appreciate the physical implications of this expression,
we can explicitly solve it by using a constant $\rho_{\rm b}(E) = \alpha$. (Note that 
this assumption is not completely unrealistic as the $t_{2g}$ bands forming
the bottom of the conduction band in many ferroelectric perovskites have
a marked 2D character; in other words, the in-plane effective mass $m_{\parallel}^*$ 
is much smaller than the out-of-plane one, $m_{\perp}^*$. Within the approximation
$m_{\perp}^*=\infty$, the constant density of states of the six-fold degenerate, 
free-electron-like 2D band is uniquely determined by $m_{\parallel}^*$.) 
This leads to
\begin{equation}
\frac{\phi_n - \phi_n^0}{\phi_n} = 
-\frac{e^2 t \lambda_{\rm eff} \alpha}{2\epsilon_0 \Omega},
\end{equation}
and with a few rearrangements to
\begin{equation}
\phi_n = \frac {\phi_n^0} {C t \lambda_{\rm eff} \tilde{\alpha} + 1},
\label{eq:spillout}
\end{equation}
where $C=e^2/2\epsilon_0$ is a constant, and $\tilde{\alpha}=\alpha/\Omega$
is the density of states per unit energy and volume of the bulk.
%
%
In spite of the drastic simplifications, Eq.~(\ref{eq:spillout}) already contains 
most of the relevant ingredients for our analysis. A few notable ones are missing
-- we shall come back to those in Sections~\ref{sec:ferrospill} and \ref{sec:bettermodel}. 
Before going into more detailed considerations, however, it is important to spell out 
the direct implications of Eq.~(\ref{eq:spillout}), which we shall be concerned with 
in the following.

First, note that all quantities appearing at the denominator at the right-hand side
of Eq.~(\ref{eq:spillout}) are positive. This means that $\phi_n$ will be negative,
and will satisfy $\phi_n^0 < \phi_n < 0$. The lower limit corresponds to the
perfect \emph{interface} screening case, $\lambda_{\rm eff}=0$. The upper limit 
corresponds to no screening, $\lambda_{\rm eff} \rightarrow \infty$. 
The situation is schematically represented in Fig.~\ref{fig:prespill}(a) and
Fig.~\ref{fig:prespill}(b). 
Given a negative $\phi_n^0$ [Fig.~\ref{fig:prespill}(a)], the charge redistribution
will induce an upward energy shift of the conduction band minimum (CBM), bringing $\phi_n$ closer to the Fermi 
level [Fig.~\ref{fig:prespill}(b)].
Second, in the limit of $t \rightarrow \infty$ (infinite thickness) $\phi_n$ 
will tend to zero from below as $\phi_n \propto -1/t$. This means that the 
self-consistent band offset $\phi_n$ is not determined by the local physical 
properties of the junction, i.e. \emph{it is no longer an interface property} 
-- the spilled-out charge will redistribute over the whole film thickness as $t$ is 
varied.
Third, the density of states of the conduction band, represented in
Eq.~(\ref{eq:spillout}) by the parameter $\alpha$, will also affect the 
value of $\phi_n$: the larger $\alpha$, the strongest the reduction in 
$\phi_n$ upon charge spill-out and electrostatic re-equilibration.
(To avoid confusion, note that in the above paragraphs, we used the 
word ``screening'' in two different contexts. By ``perfect \emph{bulk} 
screening'' we mean $\mathcal{E}_{\rm b}(D)=0$. By ``perfect 
\emph{interface} screening'' we mean $\lambda_{\rm eff}=0$.)

We can attempt a semiquantitative assessment of Eq.~(\ref{eq:spillout}) 
in a representative capacitor of thickness $t=50$ \AA{} (comparable to those that
are typically simulated within DFT). In atomic units, we use $\lambda_{\rm eff}=0.3$
(of the order of the values reported in Table~\ref{tab:deltap}), 
$C=2\pi$, and $\tilde{\alpha}=0.05$ (appropriate for
the conduction band of SrTiO$_3$, a prototypical perovskite material, with
a calculated $m_{\parallel}^*=0.77$ and $\Omega=385$ a.u.). 
We obtain 
\begin{equation}
\phi_n \sim \frac{\phi_n^0}{10}.
\end{equation}
This implies that the effect is quite strong -- even if $\phi_n^0$ is a 
rather large negative value (e.g. of the order of -1 eV), in most
practical cases the conduction charge redistribution will reduce it to 
a value that lies just below the Fermi level.
Most importantly, this implies that, when $\phi_n^0<0$, the 
\emph{physical} parameters, $\phi_n^0$ and $\lambda_{\rm eff}$, 
governing the band offset at the interface are neither accessible 
in a simulation, nor are they directly measurable in an experiment --
only $\phi_n$ might be. 
Note, however, that the ``self-consistent'' $\phi_n$ value is generally not a 
well-defined physical quantity -- this is only true within the many 
approximations used in the above derivations. In particular, we have
neglected band-bending effects: in general the electrostatic
potential will be non-uniform in the film (see Sec.~\ref{sec:bettermodel})
and $\phi_n$ will be a function of the distance from the interface. 
But even if we put this \emph{caveat} aside for a moment, the reader
should keep in mind that $\phi_n$ is determined here by \emph{space-charge} 
effects through several independent contributions. Furthermore, the film is 
no longer insulating but becomes a \emph{metal}. This is a substantial, 
qualitative departure from the physical concepts that were developed in 
the context of semiconductor/metal interfaces, and that led to the consensus 
understanding of $\phi_n$ as a genuine interface property. 

Given this situation, one needs to revisit the very foundations of the 
methodological \emph{ab-initio} approaches that have been used with
great success in the past to compute Schottky barrier heights. 
This success has critically relied on a key observation: the interface 
dipole, that one identifies with the lineup term Eq.~(\ref{eq:deltav}),
is a \emph{ground-state} property, i.e. is not directly affected
by the well-known limitations of the Kohn-Sham eigenvalue spectrum. 
This is excellent news: one can efficiently (and accurately) calculate 
$\Delta \langle V \rangle$ within DFT, and combine it with a
band-structure term ($E_{\rm V}$ or $E_{\rm C}$) calculated at a 
higher level of theory (e.g. GW); within this formally sound procedure, 
theoretical calculations have shown remarkable agreement with the 
experimental observations in the past.

In the spill-out regime (i.e. $\phi_n^0<0$) described in this Section
the above key observation \emph{no longer holds} -- the erroneous DFT 
value of $\phi_n^0$ plays a direct and dominant role in the interface 
dipole formation, as is apparent from Eq.~(\ref{eq:spillout}). 
Furthermore, as $\phi_n^0$ is systematically underestimated within
LDA or GGA, there is the concrete possibility that the spill-out
regime itself ($\phi_n^0<0$) might be an artifact of the band-gap 
problem. 
Thus, the ground-state properties of the system found in a simulation
might be \emph{qualitatively} wrong due to this issue, in loose analogy 
to, e.g., the erroneous LDA prediction of metallicity in many transition metal
compounds.
It goes without saying that the results of a simulation where 
significant spill-out of charge is found because of the mechanism
described in this Section should be regarded with great suspicion.

\subsubsection{The broken-symmetry case}
\label{sec:ferrospill}

Even if the band aligmnent is Schottky-like in the reference paraelectric
structure of the capacitor, Eq.~(\ref{eq:phid}) entails the possibility that 
it might become pathological in the ferroelectric regime (i.e. when the 
polar instability is allowed to fully relax).
Unfortunately, for this case many of the simplifying assumptions used above
are no longer valid, and for a detailed description one would need to take 
into account the more refined physical ingredients discussed in Sec.~\ref{sec:bettermodel}.
At the qualitative level, however, we can already draw some important conclusions,
as we shall briefly illustrate in the following.

\begin{figure}
    \begin{center}
    \includegraphics[width=2.8in] {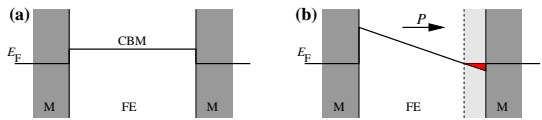}
       \caption{(Color online) (a) Paraelectric capacitor with a Schottky-like
           band alignment in the paraelectric structure. (b) 
           When the polar instability sets in, the band alignment
           becomes pathological, the conduction band is locally
           populated (red shaded area) and the film becomes partially 
           metallic (light shaded area bounded by the dashed line).
         }
       \label{fig:ferrospill}
    \end{center}
 \end{figure}

Eq.~(\ref{eq:phid}) predicts that, if $\phi_n^0$ is positive and the
capacitor is compositionally symmetric [as in Fig.~\ref{fig:ferrospill}(a)], 
at finite $D$ at most one of the two opposite interfaces will have a negative 
$\phi_n$.
%
This implies that only part of the ferroelectric film, i.e. the region
adjacent to this ``pathological'' interface, will become metallic, while
the rest of the film will stay insulating [Fig.~\ref{fig:ferrospill}(b)].
(To understand this point, note that in contrast with the previous case 
one has now a finite ``depolarizing'' electric field in the insulating 
part of the capacitor. This wedge-like potential will keep the conduction 
electrons electrostatically confined to the pathological side.)
In the insulating region, the polarization will be macroscopically constant,
as in a well-behaved capacitor [recall Eq.~(\ref{eq:maxwell})].
According to the same Eq.~(\ref{eq:maxwell}), $D(z)$ [and hence $P(z)$] 
will be non-homogeneous, with a negative slope, in the metallic region.

In this context it is worth pointing out an important physical consequence
of such a peculiar electronic ground state.
This concerns the response of the capacitor to an applied bias potential.
In well-behaved cases, the polarization of the capacitor will respond 
\emph{uniformly} to a bias, i.e. all the perovskite cells up to the electrode
interface will undergo roughly the same polar distortion.
In the present ``ferroelectric-pathological'' regime, part of the ferroelectric
film has become metallic, i.e. \emph{the metal/insulator interface has moved}
to a place that lies somewhere in the film. 
This means that, if one tries to switch the device with a potential, the electric
field won't affect the dipoles that lie closest to the pathological interface --
they will be screened by the spilled-out free charge.
A consequence is that the dipoles near a pathological interface will appear as 
if they were \emph{pinned} to a fixed distortion, that is almost insensitive 
to the electrical boundary conditions.
This pinning phenomenon has been studied in earlier theoretical works, and was 
ascribed to chemical bonding effects. In Sec.~\ref{sec:results-ferro} we shall 
substantiate with practical examples that ``dipole pinning'' is instead a direct 
consequence of the problematic band-alignment regime described here.
In Sec.~\ref{sec:discussion} we shall come back to this point and put it in the
context of the relevant literature.

\subsubsection{Towards a quantum model}
\label{sec:bettermodel}

In order to draw a closer connection between the semiclassical arguments
of the previous sections and the quantum-mechanical results that we 
present in Sections~\ref{sec:results-para} and \ref{sec:results-ferro}, 
we briefly discuss here how to improve our physical understanding of the 
charge spill-out process by lifting some of the simplifying approximations 
used so far.
As a detailed treatment goes beyond the scope of the present work, we 
shall limit ourselves to qualitative considerations.

\begin{figure}
    \begin{center}
    \includegraphics[width=2.8in] {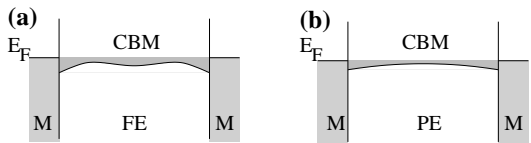}
       \caption{ 
         Schematic representation of the effects of dielectric 
         nonlinearity on the band diagram of a centrosymmetric capacitor.
         The effective potential felt by the conduction electrons is
         $-\overline{\overline{V}}_{\rm H}(z)$ (see text).
         (a): Ferroelectric material. (b): Paraelectric material.
         }
       \label{fig:nonlinear}
    \end{center}
 \end{figure}

The most drastic approximation of our model appears to be the 
assumption of perfect dielectric screening within the ferroelectric 
material, where the spill-out charge is perfectly compensated by the 
polar displacements of the lattice.
This implies that the electric field in the film vanishes, and the excess 
conduction charge can spread itself spatially at essentially no cost. 
In this scenario, the macroscopically uniform distribution of 
$\rho_{\rm free}$ postulated in Sec.~\ref{sec:paraspill} appears 
very reasonable.
In reality, the internal $\mathcal{E}$ field in the bulk ferroelectric material 
does not vanish, but is a non-linear function of $D$, that can be written by 
combining Eq.~(\ref{eq:uofd}) and Eq.~(\ref{eq:eofd}),
\begin{equation}
\mathcal{E}_{\rm b}(D) \sim \frac{1}{\Omega} \big(2 A_2 D + 4 A_4 D^3 \big)
\label{eq:eofd2}
\end{equation}
Of course, solving for the self-consistent $\rho_{\rm free}(z)$ in a non-linear 
medium would require a numerical treatment. Still, we can gain some insight 
about qualitative trends by starting, for example, from the linearly decreasing 
$D(z)$ found in the $D=0$ case of Sec.~\ref{sec:paraspill}. 
Using Eq.~(\ref{eq:eofd2}) we can write 
$\mathcal{E}(z) = \mathcal{E}_{\rm b}[D(z)]$.
The electrostatic potential is then given by integrating $\mathcal{E}(z)$. This 
essentially leads to $\overline{\overline{V}}_{\rm H}(z) = U_{\rm b}[D(z)]/Q_0$, where 
$U_{\rm b}$ is the internal electrostatic energy of Eq.~(\ref{eq:uofd}), 
and $Q_0$ is a (positive) constant with the dimension of a charge.
This means that the spatial variation in $\overline{\overline{V}}_{\rm H}(z)$ 
reflects the energy landscape of the bulk material: 
$\overline{\overline{V}}_{\rm H}(z)$  will be a double-well potential in a 
ferroelectric material ($A_2 <0$), and a single-well potential in a paraelectric 
material ($A_2 >0$).
Remarkably, the double-well potential accounts for the possibility of free-charge 
accumulation in the \emph{middle} of the centrosymmetric film [Fig.~\ref{fig:nonlinear}(a)], 
which would produce a head-to-head domain wall in the polarization $P(z)$.
Conversely, for a paraelectric material one would expect the free charge to be 
(more or less loosely) bound to the interface, and have a minimum in the middle 
of the film [Fig.~\ref{fig:nonlinear}(b)].
Of course, these considerations are valid for a centrosymmetric capacitor, and
are presented just to give the reader an idea of the physics -- in
the ferroelectric case, more complex patterns can occur and exploring them all
would require an in-depth study that is beyond the scope of this manuscript.

A second important approximation is the neglect of (i) quantum confinement effects 
beyond the simple Thomas-Fermi filling of the bulk-like density of states and (ii) 
the band-structure changes due to the polar distortions, which we briefly mentioned in
Sec.~\ref{sec:bandstr}. These will further modify the equilibrium distribution of 
the free charge, and we expect them to be important to gain a truly microscopic
understanding of the system, although not essential for the points of this work.
Remarkably, a promising model taking all these ingredients into account (dielectric 
non-linearity and band-structure effects) was recently proposed in the context of 
the (at first sight unrelated) LaAlO$_3$/SrTiO$_3$ interface.~\cite{laosto} This indicates
that the physics of a ferroelectric capacitor in the pathological band-alignment regime 
described here is essentially analogous to that of the 
``electronic reconstruction''~\cite{Okamoto-04} in oxide superlattices.
Further work to explore these interesting analogies is under way.

\subsection{Implications for the analysis of the \emph{ab-initio} results}

The above derivations show that there are two qualitatively dissimilar regimes in 
the physics of a metal/insulator interface, Ohmic-like and Schottky-like. 
During the derivation, we have evidenced some distinct physical features that
we expect to be intimately associated with the ``pathological'' Ohmic case.
As these are of central importance to help distinguish one scenario from the other,
we shall briefly summarize them in the following, mentioning also how each of these
``alarm flags'' can be detected in a first-principles simulation.

First, even after the electron re-equilibration takes place, the band edges 
cross the Fermi level of the metal, i.e. the apparent Schottky barrier is
negative. Therefore, the analysis of the local electronic structure and of
the SBH appears to be the primary tool to identify a pathological case.
However, as the ``self-consistent'' $\phi_n$ tends to stay very close to 
the Fermi level, this analysis should be performed with unusual accuracy --
techniques to do this will be discussed in Sec.~\ref{sec:schottky}.

Second, the presence of a substantial density of free charge populating the 
conduction band of the insulator is another important consequence of the
pathological regime. In Sec.~\ref{sec:deffreecharge} we illustrate how to 
rigorously define $\rho_{\rm free}$ in a ferroelectric heterostructure.

Finally, a remarkable consequence of charge spill out is the presence of
an \emph{inhomogeneous} polarization in the system. Note that this 
feature has been ascribed in earlier works to phenomena of completely
different physical origin. We shall devote special
attention in Sections~\ref{sec:results-para} and \ref{sec:results-ferro} 
to demonstrating the intimate relationship between $\rho_{\rm free}$ 
and spatial variations in $P$.

\section{Methods}
\label{sec:methods}

In this Section we spell out the practical techniques that we use to extract
the SBH from first-principles calculations, the operational definitions of
free charge and bound charge, and the methods we use to control the electrical
boundary conditions in supercell calculations. We also summarize the 
other relevant computational parameters used in Sections~\ref{sec:results-para}
and \ref{sec:results-ferro}.

\subsection{Schottky barrier estimations}
\label{sec:schottky}

First, we briefly review the methods that were used in earlier works to
compute Schottky barriers at metal/semiconductor interfaces, pointing
out advantages and limitations of each of them. Then, we
illustrate potential complications that might arise, with special
focus on ferroelectric oxide systems and the issues discussed in 
Sec.~\ref{sec:theory}.

 \subsubsection{From the local density of states}
\label{sec:ldos}

 In order to calculate the band offset at a metal/insulator
 interface, one needs to identify the location of the band edges 
 deep in the insulating region, with the Fermi level of the metal 
 taken as a reference.
 To that end, it has become common practice\cite{Peressi-98} to 
 define a spatially-resolved density of states, 
 %
 \begin{equation}
    \rho(i, E) = \sum_{n} \int_{{\rm BZ}} d{\bf{k}}
                       \left| \langle i | \psi_{n {\bf{k}}} \rangle \right|^{2}
                       \delta (E - E_{n {\bf{k}}}),
    \label{eq:ldos}
 \end{equation}
 where $|i\rangle$ is a normalized function, localized in space around the 
 region of interest.
 When $|i\rangle=|{\bf r}\rangle$ is an eigenstate of the position operator,
 the resulting $\rho({\bf r}, E)$ is commonly known as \emph{local} density of
 states (LDOS). Conversely, when $|i\rangle=|\phi_{nlm}\rangle$ is an atomic 
 orbital of specified quantum numbers $(n,l,m)$, we call 
 it instead \emph{projected} density of states (PDOS). 
 The integral is performed over the first Brillouin zone (BZ)
 of the supercell and the sum runs over all the bands $n$.
 $E_{n {\bf{k}}}$ stands for the eigenvalue of the electronic wave 
 function $\psi_{n {\bf{k}}}$. 

%
 %
 The LDOS defined in Eq. (\ref{eq:ldos}), that depends on
 the position in real space as well as
 on the energy, gives a very intuitive picture of the band offset:
%
 ``sufficiently far'' away from the interface, the LDOS converges to 
 a bulk-like curve,~\cite{Peressi-98} and in principle the location of the
 band edges (and hence the SBH) can be directly extracted by visual 
 inspection.
%
%
%
%
%
 However, several approximations are used in practice to make the
 calculation tractable, and these can introduce significant deviations 
 in the SBH computed by means of either the LDOS or PDOS.
 First, all studies are done on a finite supercell, usually with
 a symmetric capacitor geometry. This implies that the LDOS of the most 
 dispersive bands will be altered by quantum confinement effects, which 
 might produce a spurious gap opening. Also, the LDOS associated to the 
 evanescent metal-induced gap states (MIGS) might be still important at the 
 center of an insulating film that is not thick enough, thus preventing 
 an accurate identification of the band edge.
 Second, a discrete $k$-point mesh is used instead of the continuous one 
 implicitly assumed in Eq.~(\ref{eq:ldos}). Such a $k$-point mesh is generally
 optimized for efficiency, which means that high-symmetry (HS) points are
 often excluded.~\cite{k-point-sampling}
 As the edges of the valence and conduction band manifolds are usually located at 
 the HS points,\cite{Band-gap-BTO-PTO} estimating those features from the 
 calculated LDOS might lead to substantial inaccuracies. 
%
%
 For materials that display a very dispersive band structure (see e.g. 
 Ref.~\onlinecite{Delaney-10}) it is not unusual to have deviations of 
 the order of several tenths of an eV.
 Third, a fictitious electronic temperature (or Fermi surface smearing)
 is commonly used, in order to alleviate the errors introduced by the $k$-mesh
 discretization.
 This implies that the Dirac delta function in Eq.~(\ref{eq:ldos}) needs to be
 replaced by a normalized smearing function (e.g. a Gaussian) with finite
 width.
%
%
%
 This is a again potential source of inaccuracies, because the apparent edges 
 of the smeared LDOS/PDOS actually might not correspond to the \emph{physical} 
 band edges but to the (artificial) tail of the smearing function used. 

 Summarizing the above, we get to the following operational definition
 of the smeared LDOS,
 \begin{equation}
    \tilde{\rho}({\bf r}, E) = \sum_{n{\bf k}} w_{\bf k} 
                       \left|  \psi_{n {\bf{k}}}({\bf r})  \right|^{2}
                       g(E - E_{n {\bf{k}}}),
    \label{eq:discrete}
 \end{equation}
 where the BZ integral has been replaced with a sum over a discrete set
 of special points ${\bf k}$ with corresponding weights $w_{\bf k}$, and
 the Dirac delta has been replaced with a smearing function $g$.
 As will become clear shortly (a detailed analysis is provided in 
 Appendix~\ref{app:b}), it is very important to use in 
 Eq.~(\ref{eq:discrete}) a $g$ function that is minus the analytical 
 derivative of the occupation function used in the actual calculations.
 The Gaussian smearing (G) and the Fermi-Dirac (FD) smearing are by far the 
 most popular choices. These correspond to the following definitions
 of $g$,
 \begin{subequations}
    \begin{align}
 g_{\rm G}(x) & = \frac{1}{\sqrt{\pi} \sigma} e^{-x^2/\sigma^2},
       \label{eq:gg}\\
 g_{\rm FD}(x) & = \frac{\sigma^{-1}}{2+e^{x/\sigma} + e^{-x/\sigma}},
       \label{eq:gfd}
    \end{align}
 \end{subequations}
where $\sigma$ is the smearing energy used during self-consistent minimization
of the electronic ground state.
%
%
%

\subsubsection{From the electrostatic potential}

 To work around these difficulties, it is in most cases preferrable to avoid the
 direct estimation of the SBH based on the LDOS/PDOS, and use instead the indirect 
 procedure, based on the nanosmoothed electrostatic potential, 
 $\overline{\overline{V}}_{\rm H}$,
 described in Sec.~\ref{sec:metsemi}.
 The interface lineup term, $\Delta \langle V \rangle$,    
 \emph{generally} (a notable exception is the pathological spill-out regime described
 in Sec.~\ref{sec:theory} -- for further details see Sec.~\ref{sec:ldosspill}) 
 converges much faster than the LDOS/PDOS with respect to all the 
 computational parameters described above (slab thickness, $k$-mesh, Fermi surface smearing).
 The band-structure terms, $E_{\rm V}$ and $E_{\rm C}$, 
 can be then accurately and 
 economically evaluated in the bulk, without the complications
 inherent to MIGS and 
 quantum confinement effects.
 While this is in principle a very convenient and 
 robust methodological framework 
 it is, however, also prone to systematic errors. In particular, great care must be used when 
 performing the reference bulk calculations. In the vast majority of cases these
 must \emph{not} be performed on the equilibrium structure 
 of the bulk solid, but 
 will be constructed to accurately match 
 (i) the mechanical and (ii) the \emph{electrical} 
 boundary conditions of the insulating film in the supercell.
 The issue (i) is well known: in a coherent heterostructure 
 the insulating film is 
 strained to match the substrate lattice parameter, 
 and for consistency the ``bulk'' 
 calculation should be performed at the same in-plane strain. (The dependence of the
 band-structure term on the lattice strain is well known in the literature,
 and referred to as ``deformation potentials''.~\cite{Bardeen-50})
 Issue (ii) concerns ferroelectric systems, and is therefore not 
 widely appreciated
 within the semiconductor community. Whenever the symmetry of the capacitor is 
 broken and there is a net macroscopic polarization in the ferroelectric film, 
 the structural distortions may alter the band structure significantly,
 often more 
 than purely elastic effects do.~\cite{laosto}
 Note that in most capacitor calculations the film is only partially polarized 
 (i.e. it has neither the centrosymmetric non-polar structure, nor the fully 
 polarized ferroelectric structure because of the depolarizing effects described 
 in Sec.~\ref{sec:metferro}).
 The ``bulk'' reference calculation should then accurately match the polar 
 distortions of the \emph{film}, extracted in a region where the interface-related
 short-range perturbations have healed into a regular pattern.

\subsubsection{The ``best of both worlds''}
\label{sec:best}
 
 In order to minimize the drawbacks associated with either of the two methods 
 described above, we find it very convenient to combine them in the 
 following procedure.
%
%
 First, we compute the LDOS in the supercell at an atomic site (or layer)
 located far away from the interfaces, where the relaxed atomic structure 
 has converged into a regular pattern.
%
%
%
 Second, we extract the relaxed atomic coordinates from the same region 
 of the supercell, and build a periodic bulk calculation based on them, by 
 preserving identical structural distortions and strains, and by using
 an \emph{equivalent} $k$-mesh.
 (An approximation is made here, since the periodic bulk simulation is carried
 out at zero macroscopic field while the LDOS in the supercell might 
 contain the effects of a non-zero depolarizing field.
 The problem of computing the bulk layer-by-layer LDOS under a finite 
 electric field remains an open question.)
 Third, we extract the LDOS from the bulk at 
 the same atomic site or layer; we construct the bulk LDOS using 
 Eq.~(\ref{eq:discrete}) and an identical $g$ function to that used 
 in the supercell.
 Finally, we superimpose the bulk LDOS on the supercell LDOS at each layer $j$; 
 we align them by matching the sharp peaks of a selected deep semicore band, which
 are located at energies $E_{\rm sc}^{\rm supercell}(j)$ and 
 $E_{\rm sc}^{\rm bulk}$.
 The deep semicore states are insensitive to the chemical environment 
 and have negligible band dispersion; this means that they provide an excellent,
 spatially localized reference energy for the estimation of the lineup term.

 At this point, we look at either LDOS curve in the vicinity of the Fermi 
 level. If it is non-zero we are probably facing a pathological spill-out
 case (see the following Section).
 If it is zero, then we can go one step further and accurately estimate 
 the positions of the local band edges. 
 To this end, we compute from the bulk calculation
 $E_{\rm C}^{\rm bulk}$ and $E_{\rm V}^{\rm bulk}$, together with $E_{\rm sc}^{\rm bulk}$. 
 (A further non-selfconsistent run might
 be needed if the original $k$-mesh did not include the HS $k$-points 
 where the band edges are located.)
 Finally, assuming that $E_{\rm sc}^{\rm supercell}(j)$ are all referred
 to an energy zero corresponding to the self-consistent Fermi level of
 the supercell, we define the local position of the band edges as
\begin{equation}
 E_{\rm C,V}^{\rm supercell}(j) = E_{\rm sc}^{\rm supercell}(j) - E_{\rm sc}^{\rm bulk} + 
E_{\rm C,V}^{\rm bulk}
\label{eq:semicore}
\end{equation}
 This procedure avoids the (often inaccurate) estimate of the band edges
 based on the tails of the smeared LDOS, and at the same time preserves the
 advantages of the ``lineup + band structure'' technique. 
 In principle, the latter method should accurately match the results of 
 Eq.~(\ref{eq:semicore}), except for quantum confinement effects in the 
 metallic slab used to represent the semi-infinite electrode, as 
 discussed in Ref.~\onlinecite{Fall-99}.
%

 Note that this technique is not only useful to detect pathological 
 band alignments and extract accurate band offsets in the non-pathological 
 cases.
 Given that we are superimposing two LDOS calculated with  
 identical computational parameters and structures, their direct comparison 
 can be very insightful.
 Most importantly, one expects all the features to 
 closely match \emph{unless} there 
 are MIGS or confinement effects. Therefore, one has also a powerful tool to 
 directly assess the impact of the latter physical ingredients in the supercell 
 electronic structure. This procedure, therefore, yields far
 more physical information 
 than the separate use of either the PDOS/LDOS or the nanosmoothing method.


\subsubsection{The pathological regime}
\label{sec:ldosspill}

 In the pathological regime described in Sec.~\ref{sec:theory}, many of
 the conditions that formally justify application of the above methods to 
 the estimation of the SBH break down.
 First, the presence of a non-uniform electric displacement $D(z)$ 
 implies that the polar distortions are also non-uniform, and they may
 \emph{not} converge to a regular bulk-like pattern anywhere in the
 film.
 Second, electrostatic and exchange and correlation effects due to the 
 partial filling of the conduction band imply that the band structure 
 may significantly depart from what one computes in the \emph{insulating}
 bulk (note that this is distinct from the effect of the structural
 distortions discussed in the previous Section).
 Third, the usual assumption of fast convergence of the interface dipole
 with respect to slab thickness, $k$-mesh resolution and smearing energy 
 also breaks down, as the conduction band DOS (which converges slowly with
 respect to these parameters) is now \emph{directly} involved in the 
 electrostatic re-equilibration process.
 Based on this, the reader should keep in mind that there is an intrinsic 
 arbitrariness, of \emph{physical} more than methodological nature, in 
 the definition of the band edges in spill-out cases. 
 This arbitrariness reflects itself in the fact, already pointed out in 
 Sec.~\ref{sec:theory}, that the band alignment at a pathological 
 interface is no longer a well-defined interface property, nor is it directly 
 measurable in an experiment.
 The position of the bands is essentially the result of a complex electron
 redistribution process that may occur on a scale that is almost macroscopic, 
 and is driven by different factors than those usually involved in the SBH
 formation.

 Of course, by using all the precautions that are valid at well-behaved 
 interfaces one might still gain some qualitative insight into the local 
 electronic properties of the system. However, the data must be interpreted 
 with some caution, and it is most appropriate to combine the analysis with 
 other post-processing tools before drawing any conclusion.
 We shall discuss some of these further analysis tools in the
 following Sections.
 
\subsection{Electrical analysis of the charge spill-out}

 In this Section we introduce the methodological tools that we
 use to analyze in practice the spill-out regime, in light of
 the theory developed in Sec.~\ref{sec:theory}.
%
%
 In particular, we illustrate how to rigorously define 
 the ``local electric displacement'' $D(z)$ and the 
 ``conduction charge'' $\rho_{\rm free}$.
 To evaluate the former, we discuss two approaches. The
 first one is based on a Wannier decomposition of the bound
 charges. The second one is an approximate formula in terms of
 the ionic distortions and the Born effective charges (BEC).
 This simplified formula is very practical for a quick analysis,
 but is generally affected by systematic errors.
 We address this issue by proposing a simple correction that
 significantly improves the accuracy of the BEC estimate.

\subsubsection{Definition of bound charge and conduction charge}
\label{sec:deffreecharge}

 In a typical metal, it is difficult to rigorously identify
 conduction electrons and bound charges, as usually the respective
 energy bands intersect each other in at least some regions of the
 Brillouin zone.
 (This is true, for example, in all transition metals, where the
 delocalized $sp$ bands cross the more localized $d$ bands.)
 By contrast, in all perovskite materials considered here,
 even upon charge spill out and metallization, a well-defined
 energy gap persists between the bound electrons and the
 partially filled conduction bands.
 Therefore, it is straightforward to separate the
 two types of charge densities, free and bound, simply
 by integrating the local density of states, defined in Eq. (\ref{eq:discrete}), 
 over two distinct energy windows.
 For example, for the conduction charge $\rho_{\rm free}$ we have
 \begin{equation}
    \rho_{\rm free} (\mathbf{r}) = \int_{E_0}^{E_{\rm F}} 
    \tilde{\rho}(\mathbf{r},E) dE = \sum_{E_{n{\bf k}}>E_0} w_{\bf k} f_{n{\bf k}} 
                       \left|  \psi_{n {\bf{k}}}({\bf r})  \right|^{2},
    \label{eq:rhofree}
 \end{equation}
 where $E_0$ is an energy corresponding to the center
 of the gap between valence and conduction band, $\tilde{\rho}$ 
 is the smeared density of states of Eq.~(\ref{eq:discrete}),
 $f_{n{\bf k}}$ are the occupation numbers and the sum is restricted 
 to the states with eigenvalue $E_{n{\bf k}}$ higher than $E_0$.
 [Note that Eq.~(\ref{eq:rhofree}) only holds if the $g$-smearing 
 of $\tilde{\rho}$ is compatible with the definition of $f_{n{\bf k}}$,
 see the last paragraph of Sec.~\ref{sec:ldos} and Appendix~\ref{app:b}.]
 Since we are working with layered systems that are perfectly
 periodic in plane, we will be mostly concerned with the planar
 average of $\rho_{\rm free}$,

 \begin{equation}
    \overline{\rho}_{\rm free} (z) = \frac{1}{S} \int_S
    \rho_{\rm free}(\mathbf{r}) dxdy,
    \label{eq:planaverhofree}
 \end{equation}

 \noindent where $S$ is the area of the interface unit cell.
 In some cases, it is also useful to consider the
 nanosmoothed function,~\cite{Junquera-07} which we indicate by
 a double bar symbol, $\overline{ \overline{\rho}}_{\rm free} (z)$.

 Concerning the bound charges, we shall approximate the local electric 
 displacement $D(z)$ with the local polarization $P(z)$. This is an
 excellent approximation in many ferroelectric materials, where $P$
 is of the order of 0.1-1 C/m$^{-2}$ and $D-P = \epsilon_0 \mathcal{E}$
 is typically much smaller than 10$^{-3}$ C/m$^{-2}$. (The largest 
 electric fields $\mathcal{E}$ that can be applied without dielectric
 breakdown~\cite{Grigoriev} are of the order of 0.1 GV/m.) Thus, 
 assuming $D(z)\sim P(z)$ entails errors of 1\% or less, which we 
 consider negligible for the purposes of our discussion.
 Techniques to extract $P(z)$ from a supercell calculation are described 
 in the following sections.

\subsubsection{Local polarization via Wannier functions}
\label{sec:localpolwan}

 A very useful tool to describe the local polarization properties
 of layered oxide superlattice are the ``layer polarizations''
 introduced by Wu {\it et al.}~\cite{Xifan_lp}
 First, we transform the electronic ground state into a set of
 ``hermaphrodite'' Wannier orbitals~\cite{Xifan_lp,eamonn_kn}
 by means of the parallel-transport~\cite{Marzari/Vanderbilt:1997}
 procedure.
 Note that we restrict the parallel-transport procedure only to
 the orbitals that we consider as ``bound charge'', i.e. those
 with an energy eigenvalue lower than $E_{\rm 0}$.
 Then, we group the Wannier centers and the ion cores into individual
 oxide layers, and define the dipole density of layer $j$ as

 \begin{equation}
    p_j = \frac{1}{S} \Big(
          \sum_{\alpha \in j} Z_\alpha R_{\alpha z} -
          2e \sum_{i \in j} z_i \Big)  ,
 \end{equation}

 \noindent where $Z_\alpha$ is now the \emph{bare} valence charge 
 of the atom $\alpha$, whose position along $z$ is $R_{\alpha z}$,
 and $z_i$ is the location of the Wannier orbital $i$.

 Note that individual oxide layers in II-IV perovskites like BaTiO$_3$ or
 PbTiO$_3$ are charge-neutral and the $p_j$ are well-defined; however, in
 I-V perovskites like KNbO$_3$, individual layers are charged, and the $p_j$
 become meaningless as they are origin dependent.
 To circumvent this problem, one can either combine the layers two by two
 as was done in Ref.~\onlinecite{Neaton-03}, or perform some averaging with
 the neighboring layers, as for example in Ref.~\onlinecite{eamonn_kn}.
 It is important to keep in mind that, depending on the specific averaging
 procedure, one might end up with the \emph{formal} or with the
 \emph{effective} local polarization;\cite{Stengel-09.1}
 in this work we find it
 more convenient to work with the latter. As we do not need, for the purpose
 of our discussion, to resolve $P$ into contributions from individual AO and
 BO$_2$ oxide layers, at variance with Ref.~\onlinecite{eamonn_kn} we perform a
 simple average

 \jjm{Should we use at ``variance'' here?. In Ref.~\onlinecite{eamonn_kn}
      also perform averages with neighbour layers.}
 \msm{Not really an average. They keep a layer-by-layer spatial resolution, 
      and correct for the excess/defect charge in some slightly complicated
      way. Here it is just a simple average of everything... you cannot
      distinguish AO from BO$_2$ layers any more.}
 \begin{equation}
    \bar p_j = \frac{1}{4} p_{j-1} + \frac{1}{2} p_j + \frac{1}{4} p_{j+1}.
 \end{equation}

 \noindent We then define the local \emph{polarization} by scaling this surface
 dipole density by the average out-of-plane lattice parameter, $c$, of the
 oxide film, and by taking into account that every individual oxide layer
 occupies only half the cell. We thus define the local polarization as
 
 \begin{equation}
    P_j = \frac{2}{c} \bar p_j.
 \end{equation}

 The local polarization $P_j$ is, of course, a discrete set of
 values, but we can think of it as a continuous function of the
 $z$ coordinate, $P(z)$, which is sampled at the oxide plane locations.
 In the remainder of this work, we will write $P_j$
 or $P(z)$ depending on the context, but the reader should bear in mind
 that these two notations refer to the same object.

\subsubsection{Approximate formula via Born effective charges}
\label{sec:localpolborn}

 While the above definition of $P_j$ in terms of Wannier functions is
 accurate and rigorous, it is not immediately available in most
 electronic structure codes.
%
%
%
 An approximate estimate of the local polarization can be simply 
 inferred from the bulk Born effective charges $Z^{\ast}_\alpha$ and the
 local atomic displacements.
%
 Analogously to the above formulation, we can write the $Z^{\ast}_\alpha$-based 
 approximate layer dipole density, $p^Z_j$, as
 \begin{equation}
    p^Z_j = \frac{1}{S}
    \sum_{\alpha \in j} Z^{\ast}_\alpha R_{\alpha z} ,
    \label{eq:localpolbec1}
 \end{equation}

 \noindent where $Z^{\ast}_\alpha$ is now the bulk Born effective 
 charge associated with the atom $\alpha$. 
 \jjm{Explain the meaning of the superscript $Z$ here.}
 \msm{Slightly modified the sentence before the equation. 
      Is it clear now?}
 Again, $p^Z_j$ are ill-defined in perovskite materials, as typically
 individual oxide layers do not satisfy the acoustic sum rule separately.
 To address this issue, we perform an analogous averaging procedure
 and define

 \begin{equation}
   \bar p^Z_j = \frac{1}{4} p_{j-1} + \frac{1}{2} p_j + \frac{1}{4} p_{j+1}.
    \label{eq:localpolbec2}
 \end{equation}

 The approximate local polarization then immediately follows,

 \begin{equation}
    P^Z_j = \frac{2}{c} \bar p^Z_j.
    \label{eq:localpolbec3}
 \end{equation}
 \jjm{I guess a superscript $Z$ is missing in $p_j$ here.}
 \msm{Fixed.}

 Such an approximation provides an exact estimate, in the linear limit,
 of the polarization induced by a \emph{small} polar distortion under
 \emph{short-circuit} electrical boundary conditions, i.e. assuming
 that the macroscopic electric field vanishes throughout the
 structural transformation.
 Neither of these conditions is respected in a ferroelectric capacitor,
 where the polar distortion is generally large (close to the spontaneous
 polarization of the ferroelectric insulator), and where there is generally
 an imperfect screening regime, with a macroscopic ``depolarizing
 field''.~\cite{Junquera/Ghosez:2003}
 We investigate both issues in the Appendix, where we find that a
 simple scaling factor corrects, to a large extent, the discrepancy
 between $P_j$ and $P^Z_j$. In particular, we write the ``corrected''
 $\tilde P^Z_j$ as

 \begin{equation}
    \tilde P^Z_j = \left( 1 + \frac{\chi_\infty}{\chi_{\rm ION}}\right) P^Z_j,
    \label{eqzstar}
 \end{equation}

 \noindent where $\chi_\infty$ and $\chi_{\rm ION}$ are, respectively, the
 electronic and ionic susceptibilities of the bulk material in the
 centrosymmetric reference structure, calculated at the same in-plane
 strain as the capacitor heterostructure.
 Note that for a ferroelectric material in the centrosymmetric reference
 structure, $\chi_{\rm ION}$ is \emph{negative}, which is a consequence
 of the polar unstable mode in the phonon spectrum.
 This means that the scaling factor will be smaller than 1 ($\sim$0.9
 for the materials considered in this work).
 Practical methods to calculate $\chi_\infty$ and $\chi_{\rm ION}$ are
 reported in the Appendix.

\subsection{Constrained-$D$ calculations}
\label{sec:constrainedd}

 In Sec.~\ref{sec:theory} we have shown that a pathological spill-out 
 regime can be triggered by the ferroelectric displacement $D$ of the
 film, as the band offset generally strongly depends on $D$.
 It is therefore important, in order to perform the analysis described in the
 previous sections, to calculate the electronic and structural ground state 
 of a metal/ferroelectric interface at different values of $D$. 
 To this end, we can use two different approaches in 
 first-principles calculations.
 The first, and more ``traditional'' approach, involves the 
 construction of capacitor of varying thicknesses $t$, 
 and the relaxation of the corresponding ferroelectric ground states 
 within short-circuit boundary conditions.
 Due to the interface-related depolarizing effects mentioned in
 Sec.~\ref{sec:theory} (these are strongest in
 thinner films and tend to reduce $P$ from the bulk value $P_{\rm s}$),
 the polarization will increase from $P=0$ (for $t < t_{\rm crit}$,
 where $t_{\rm crit}$ is the ``critical thickness''~\cite{nature_mat,Junquera/Ghosez:2003})
 to $P \sim P_{\rm s}$, in the limit of very large thicknesses.
%
%
 This might be cumbersome in practice: thicker capacitor heterostructures
 imply a substantial computational cost, due to the larger size of the
 system; this severely limits the range of $P$ values that can be studied
 within short-circuit boundary conditions.

 An alternative, more efficient methodology to explore the electrical
 properties of the interface as a function of polarization, is to use
 the recently developed techniques to constrain the macroscopic electric
 displacement to a fixed value.~\cite{fixedd,Stengel-09.2}
 With this method, one is able, in principle, to access at the same
 computational cost the structural and electronic polarization of the
 capacitor for an arbitrary polarization state.
 In the specific context of the present work, however, there are two
 drawbacks related to the use of the constrained-$D$ method as 
 implemented in Refs.~\onlinecite{fixedd} and \onlinecite{Stengel-09.2}.
 First, fixed-$D$ strategies make use of applied electric
 fields to control the polarization of the system. This is a
 problem here, where the metallicity associated with the space charge
 which populates the ferroelectric film makes such a
 solution problematic. (If a capacitor becomes metallic, it is a conductor
 and no metastable polarized state can be defined at any given bias.)
 Second, our philosophy in this work is to adopt ``standard''
 computational techniques, i.e. those that are in principle
 available in any standard electronic structure package.

 To this end, we introduce here an alternative way of performing
 constrained-$D$ calculations for a metal-insulator interface, which
 does not rely on the direct application of macroscopic electric fields
 or on the calculation of the macroscopic Berry-phase polarization.
 We adopt a vacuum/ferroelectric/metal geometry.
%
 To induce a given value of the polarization in the ferroelectric
 film, we introduce a layer of bound charges ($Q$ per surface unit
 cell $S$) at its free surface.
 If we do so in such a way that the surface region remains \emph{locally insulating},
 at electrostatic equilibrium, the difference in the macroscopic
 displacement $D$ on the left and on the right side of the surface will
 exactly correspond to the additional surface charge density $Q/S$.
 By applying a dipole correction in the vacuum region, we ensure that
 $D=0$ in the region near the surface on the vacuum side; then on the
 insulator side we have exactly
 \begin{equation}
    D=\frac{Q}{S}.
\label{eq:sigma}
 \end{equation}
 In practice, the additional charge density is introduced by
 substituting a cation at the ferroelectric surface by a
 fictitious cation of different formal valence.
%
%
 As we are interested in exploring intermediate values of $D$,
 we use the virtual crystal approximation to effectively induce
 a fractional nuclear charge.

 The reader might have noted that this method to control $D$ is
 just a generalization of Eq.~(\ref{eq:maxwell}) to consider other
 forms of ``external'' charge that are not ``free'' in nature.
 Indeed, in the most general case, one can state
\begin{equation}
 \nabla \cdot {\bf D}({\bf r}) = \rho_{\rm ext}({\bf r}),
\label{eq:rhoext}
\end{equation}
 where ${\bf D}$ encompasses all bound-charge effects that can
 be referred to the properties of a periodically repeated primitive
 bulk unit, and $\rho_{\rm ext}$ contains all the rest (e.g. delta-doping
 layers, metallic free charges, charged adsorbates, variations in the 
 local stoichiometry, etc.).
 In Eq.~(\ref{eq:sigma}) we simply applied Eq.~(\ref{eq:rhoext}) to
 the vacuum/ferroelectric interface, where the ``bound'' nature of the
 external charge allows us to control it as an external parameter.

\subsection{Computational parameters}

 To demonstrate the generality of our arguments,
 which are largely independent of the fine details
 of the calculation (except for the choice of the density functional),
 we use two different DFT-based electronic structure codes,
 {\sc Lautrec} and {\sc Siesta}.~\cite{Soler-02}
 In both cases, the interfaces were simulated by using
 a supercell approximation with periodic boundary
 conditions.~\cite{Payne-92}
 A $(1 \times 1)$ periodicity of the supercell 
 perpendicular to the interface is assumed. 
 This inhibits the appearance
 of ferroelectric domains and/or tiltings and rotations of the O octahedra.
 A reference ionic configuration was defined by piling up
 $m$ unit cells of the perovskite oxide
 (PbTiO$_{3}$, BaTiO$_{3}$, 
 or KNbO$_{3}$), and
 $n$ unit cells of the metal electrode
 (either a conductive oxide, SrRuO$_{3}$, or a transition metal, Pt).
%
 In order to simulate the effect of the mechanical boundary conditions
 due to the strain imposed by the substrate, the in-plane
 lattice constant was fixed to the theoretical equilibrium lattice
 constant of bulk SrTiO$_{3}$ ($a_{0}$ = 3.85 \AA\ for {\sc Lautrec}
 and $a_{0}$ = 3.874 \AA\ for {\sc Siesta}).
%

 To simulate the capacitors in an unpolarized configuration in
  Sec. \ref{sec:results-para}, we imposed a mirror symmetry plane at
 the central BO$_{2}$ layer, where B stands for Ti or Nb, and relaxed
 the resulting tetragonal supercells within $P4/mmm$ symmetry. 
 For the ferroelectric capacitors described in Sec.~\ref{sec:results-ferro}
 a second minimization was carried out, with the constraint of the
 mirror symmetry plane lifted.
 Tolerances for the forces and stresses are 0.01 eV/\AA\ and 
 0.0001 eV/\AA$^{3}$, respectively. 
 Other computational parameters, specific to each code,
 are summarized below.

\subsubsection{{\sc Lautrec}}

 Calculations in Sec. \ref{sec:unpolarpath} and
 \ref{sec:opencircuit} were performed with {\sc Lautrec}, an ``in-house''
 plane-wave code based on the projector-augmented wave
 method.~\cite{Bloechl:1994}
 We used a plane-wave cutoff of 40 Ry and a $ 6 \times 6 \times 1 $
 Monkhorst-Pack \cite{Monkhorst-76,Moreno-92} mesh. As the systems
 considered here are metallic, we adopted a Gaussian smearing of
 0.15 eV to perform the Brillouin-zone integrations.

\subsubsection{{\sc Siesta}}

 Computations in Sec. \ref{sec:unpolarnonpath} and
 \ref{sec:shortcircuit} on
 short-circuited SrRuO$_{3}$/PbTiO$_{3}$
 and SrRuO$_{3}$/BaTiO$_{3}$ capacitors 
 were performed within
 a numerical atomic orbital method,
 as implemented in the {\sc Siesta} code.~\cite{Soler-02}
 Core electrons were replaced by fully-separable \cite{Kleinman-82}
 norm-conserving pseudopotentials, generated following the recipe given
 by Troullier and Martins.~\cite{Troullier-91}
 Further details on the pseudopotentials and basis sets
 can be found in Ref.~\onlinecite{Junquera-03.2}.

 A $ 6 \times 6 \times 1 $ Monkhorst-Pack \cite{Monkhorst-76,Moreno-92}
 mesh was used for the sampling of the reciprocal space.
 A Fermi-Dirac distribution was chosen for the occupation of the
 one-particle Kohn-Sham electronic eigenstates, with a smearing temperature
 of 0.075 eV (870 K).
 The electronic density, Hartree, and exchange-correlation potentials,
 as well as the corresponding matrix elements between the basis orbitals, were
 computed on a uniform real space grid, with an
 equivalent plane-wave cutoff of 400 Ry in the representation of the
 charge density.


\section{Results: Paraelectric capacitors}
\label{sec:results-para}

\subsection{Non-pathological cases}
\label{sec:unpolarnonpath}

\begin{figure}
    \begin{center}
    \includegraphics[width=\columnwidth,clip]{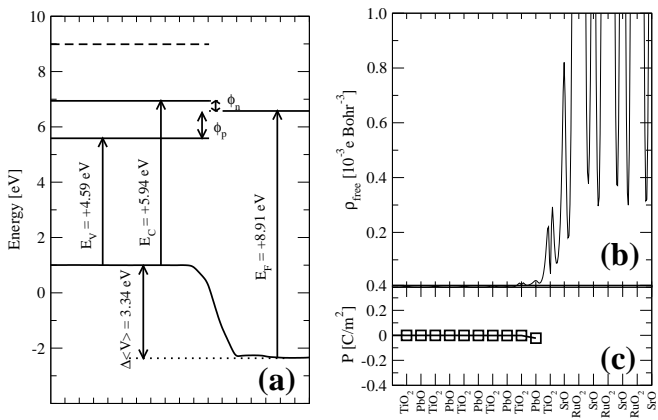}
       \caption{(a) Schematic representation of $\phi_{n}$ and $\phi_{p}$ in
                 an unpolarized SrRuO$_{3}$/PbTiO$_{3}$/SrRuO$_{3}$ capacitor.
                 $E_{\rm V}$, $E_{\rm C}$, $E_{\rm F}$ and 
                 $\Delta \langle V \rangle$ were defined in
                 Sec.~\ref{sec:metsemi}. The calculated values are also
                 indicated in the Figure. The black solid line represents 
                 $-\overline{\overline{V}}_{\rm H}(z)$.
                 The dashed line represents the hypothetical position of the 
                 CBM if $E_{\rm C}$ were shifted to reproduce the 
                 experimental band gap.
                 (b) Profile of $\bar{\rho}_{\rm free}$ as defined
                 in Eq. (\ref{eq:planaverhofree}).
                 (c) Profile of the layer-by-layer polarization $P^Z_j$.
                 The size of the capacitor corresponds to $n$ = 5.5 unit cells
                 of SrRuO$_{3}$ and $m$ = 12.5 unit cells of PbTiO$_{3}$. Only 
                 the top half of the symmetric supercell is shown.
         }
       \label{fig:unpolldapto}
    \end{center}
 \end{figure}

 In the centrosymmetric unpolarized reference structure,
 some metal/ferroelectric interfaces
 such as BaTiO$_{3}$/SrRuO$_{3}$ or PbTiO$_{3}$/SrRuO$_{3}$
 are ``well-behaved'' within LDA. [We focus here on the TiO$_{2}$/SrO termination -- the properties
 of the alternative (Ba,Pb)O/RuO$_2$ termination might differ.] 
 This conclusion emerges from the analysis shown in Fig. \ref{fig:unpolldapto} 
 for the PbTiO$_{3}$-based capacitor; qualitatively similar results, 
 not shown here, are obtained for the BaTiO$_{3}$-based capacitor.
 Figure \ref{fig:unpolldapto}(a) represents
 schematically the Schottky barriers for electrons ($\phi_{n}$) and
 holes ($\phi_{p}$) at the ferroelectric/metal interfaces, computed
 using the nanosmoothed electrostatic potential method described in
 Sec. \ref{sec:metsemi}.
 The bottom of the conduction band of the ferroelectric lies above the
 Fermi level of the metal ($\phi_{n}$ amounts to 0.38 eV for the
 PbTiO$_{3}$-based capacitor, and only to 0.19 eV in the 
 BaTiO$_{3}$-based case).
 Note that, if the experimental band gap could be reproduced
 in our simulations, $\phi_{n}$ would be much larger [dashed 
 lines in Fig. \ref{fig:unpolldapto}(a); we have taken the 
 experimental indirect gap of the cubic phase of PbTiO$_{3}$, 
 3.40 eV~\cite{Peng-92} and assumed that the quasiparticle 
 correction on the valence band edge is negligible].
 The results summarized in Table~\ref{table:schottkybarriers}
 indicate that, in all the cases discussed here, different 
 methodologies yield Schottky barrier values that are consistent 
 within a few hundredths of an eV.
 The flatness of the profile of the nanosmoothed electrostatic
 potential at the central layers of PbTiO$_3$ 
 confirms the absence of any macroscopic electric field,
 as expected from a locally charge-neutral
 and centrosymmetric system.

 \begin{table}
    \caption[ ]{ LDA values of $\phi_{n}$ and $\phi_{p}$,
                 obtained with two different methods: using the
                 decomposition into $E_{\rm V,C}$ and 
                 $\Delta \langle V \rangle$ (BS + Lineup),
                 or using the method of Sec.~\ref{sec:best} (Semicore).
                 In the ``Semicore'' case we used the sharp Ti(3s)-derived 
                 peak of the LDOS (extracted from the central TiO$_{2}$ layer
                 of the capacitor) to align the energies of the bulk band edges
                 with the supercell Fermi level.

               }
    \begin{center}
       \begin{tabular}{ccc}
          \hline
          \hline
          Capacitor                             &
          BS + Lineup                           &
          Semicore                                  \\
	   \hline
          SrRuO$_{3}$/PbTiO$_{3}$/SrRuO$_{3}$   &
                                                &
                                                \\
          $\phi_{p}$ (eV)                       &
          0.97                                  &
          0.99                                  \\
          $\phi_{n}$  (eV)                      &
          0.38                                  &
          0.37                                  \\
          SrRuO$_{3}$/BaTiO$_{3}$/SrRuO$_{3}$   &
                                                &
                                                \\
          $\phi_{p}$  (eV)                      &
          1.39                                  &
          1.41                                  \\
          $\phi_{n}$  (eV)                      &
          0.19                                  &
          0.23                                  \\
          \hline
          \hline
       \end{tabular}
    \end{center}
    \label{table:schottkybarriers}
 \end{table}

 \begin{figure}
    \begin{center}
    \includegraphics[width=\columnwidth,clip] {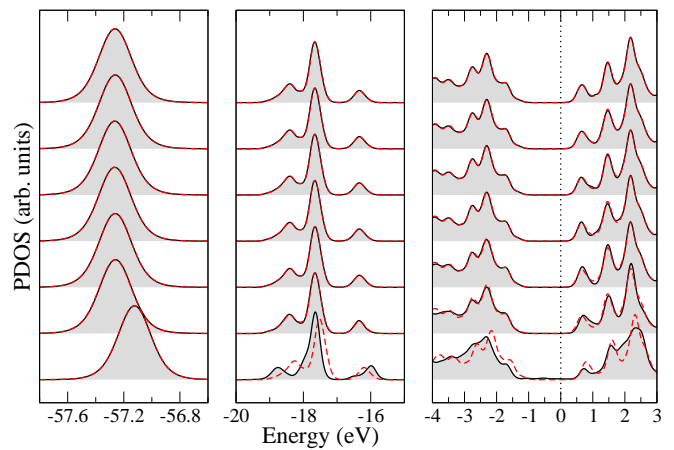}
       \caption{ (Color online) 
                 PDOS of the inequivalent TiO$_2$ layers in the 
                 unpolarized PbTiO$_3$/SrRuO$_3$ 
                 capacitor (solid curves with gray shading). 
                 The bottom curve lies next to the 
                 electrode, the top one lies in the center of the PbTiO$_3$
                 film. Only the PDOS on half of the symmetric supercell are
                 shown. The bulk PDOS
                 curves (red dashed) are aligned to match 
                 the Ti(3s) peak at $E\sim -57$ eV.
                 The Fermi level is located at zero energy.
         }
       \label{fig:paraldos}

    \end{center}
 \end{figure}

 Figure \ref{fig:unpolldapto}(b)
 displays $\bar{\rho}_{\rm free}(z)$, as defined in 
 Sec.~\ref{sec:deffreecharge}.
 As expected, $\bar{\rho}_{\rm free}(z)$ has a rapid decay in 
 the insulating layer, consistent with the evanescent character
 of the metallic states (MIGS): these cannot propagate in the 
 insulator as their energy eigenvalues fall within the forbidden
 band gap. 
%
%
%
 Fig. \ref{fig:unpolldapto}(c) shows the layer-by-layer polarization,
 $P^Z_j$, 
 computed using Eqs.~(\ref{eq:localpolbec1})-(\ref{eq:localpolbec3}).
 Consistent with the absence of space charge, the $P^Z_j$ profile is 
 remarkably flat. Due to the imposed mirror-symmetry constraint,
 $P^Z_j$ also vanishes inside the ferroelectric material.

 Fig. \ref{fig:paraldos} shows the layer-resolved PDOS of the
 Ti(3s) semicore peaks, the O(2s) peak, the upper valence band 
 and the lower conduction band (black curves, shaded in gray).
 On top of the heterostructure PDOS we superimpose the bulk PDOS,
 calculated with an equivalent $k$-point sampling and aligned with
 the Ti(3s) peak (dashed red curves). 
 Note that all PDOS curves were calculated using Eq.~(\ref{eq:discrete}),
 and the smearing function $g_{\rm FD}$ of Eq.~(\ref{eq:gfd}) with $\sigma=0.075$
 eV, consistent with the parameters used in the calculation.
 The PDOS of the conduction and valence bands converges fairly quickly to the 
 bulk curve when moving away from the interface -- they are practically 
 indistinguishable already at the fourth layer.
 The estimated energy locations of the conduction and valence bands converge
 even faster [these are directly related to the shifts of the Ti(3s) state,
 which are less affected by quantum confinement effects].
 All curves except those adjacent to the electrode interface vanish
 at the Fermi level, confirming the absence of charge spill-out in
 this system.

 As a summary of this Section we can conclude that,
 when a centrosymmetric unpolarized interface is non-pathological
 in the sense that the bottom of the conduction band of the
 ferroelectric is above the Fermi level of the metal,
 (i) the free charge, as defined in Sec. \ref{sec:deffreecharge},
 vanishes due to the absence of charge spill-out; (ii) 
 the local polarization profile (Sec. \ref{sec:localpolborn}) is 
 perfectly flat as the interface-induced \emph{polar} lattice distortions
 heal rapidly (within the first unit cell); (iii) the
 LDOS/PDOS vanishes at the Fermi level, except for one or two
 interface layers where the signatures of the MIGS might be still 
 present (they are barely detectable in the curves of 
 Fig.~\ref{fig:paraldos}).

\subsection{Pathological cases}
\label{sec:unpolarpath}

 We analyze now two examples of capacitors that are characterized by a
 pathological band alignment already in their centrosymmetric reference
 structure: NbO$_2$-terminated KNbO$_3$/SrRuO$_3$, and
 TiO$_2$-terminated BaTiO$_3$/Pt.
 This choice of materials is motivated by the fact that there exist recent
 theoretical works on these systems,~\cite{Duan-06,Velev-07} where the
 consequences of the pathological band alignment were neglected. 

\subsubsection{KNbO$_3$/SrRuO$_3$}
\label{sec:knosro}

 \begin{figure}
    \begin{center}
       \includegraphics[width=0.45\textwidth,clip]{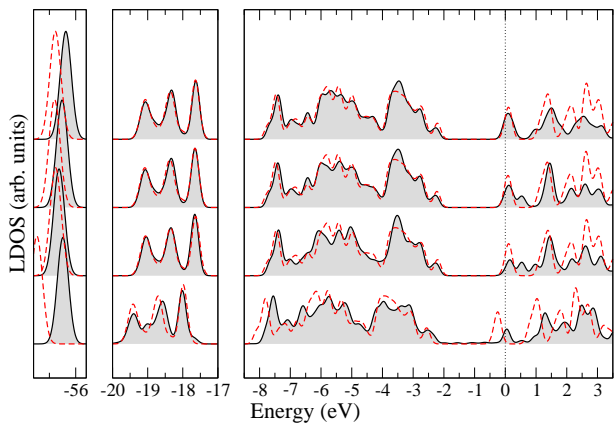}
    \end{center}
    \caption{ (Color online) LDOS integrated over the NbO$_{2}$ layers of the
              KNbO$_{3}$/SrRuO$_{3}$ heterostructure 
              (solid curves with gray shade). 
              The bottom curve lies next to the electrode, the top one lies 
              in the middle of the KNbO$_{3}$ film. Only the PDOS on half
              of the symmetric supercell are shown. The bulk LDOS (red dashed 
              curves) are aligned to match the valence and conduction band 
              edges. The Fermi level is located at zero energy.}
    \label{fig:knoldos}
 \end{figure}

 We construct a heterostructure consisting of $m$=6.5 KNbO$_{3}$ 
 unit cells and $n$=7.5
 SrRuO$_{3}$ cells, for a total of 14 perovskite units; 
 we use symmetrical NbO$_2$ (SrO) terminations of the KNbO$_{3}$ 
 (SrRuO$_{3}$) film.
%
%
 After full relaxation with a mirror symmetry constraint at the 
 central NbO$_{2}$ layer, we perform the analysis of the LDOS, the 
 conduction charge and the local polarization as explained in 
 Sec.~\ref{sec:methods}.
 In Fig.~\ref{fig:knoldos} we show the local density of states integrated over
 the NbO$_{2}$ layers (the bottom one is adjacent to the electrode interface,
 the top one lies on the mirror plane in the middle of the film). 
%
%
 The unphysical Ohmic band alignment is evident from the location of 
 the conduction band bottom -- the whole film is clearly \emph{metallic}. 
 This points to the pathological situation that is sketched in 
 Fig. \ref{fig:prespill}. Note that the LDOS does not converge to the
 bulk curve anywhere in the heterostructure. There are non-trivial shifts 
 of all peaks that make it difficult to identify a well-defined alignment 
 with the bulk curves. In Fig.~\ref{fig:knoldos} we choose to align the
 O(2s)-derived feature at $E\sim-19$ eV. In this specific system, aligning
 the O(2s) peaks appears to yield a reasonably good match of the conduction 
 and valence band edges (the most relevant features from a physical point 
 of view); this, however, leads to a marked mismatch, e.g. in the position 
 of the semicore Nb(4s) state. We show in the following that these effects 
 stem from a number of (rather dramatic) electrostatic and structural 
 perturbations acting on the KNbO$_3$ film, which are a direct consequence 
 of the pathological band alignment.

 \begin{figure}
    \begin{center}
       \includegraphics[width=0.35\textwidth,clip]{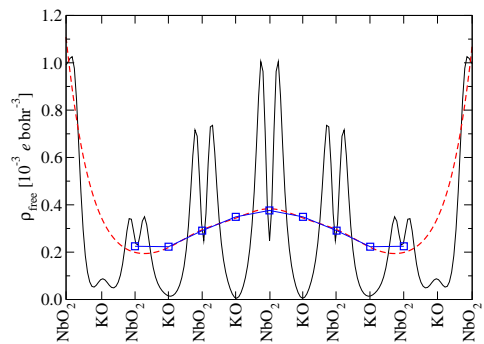}
    \end{center}
    \caption{ (Color online) Calculated free charge for paraelectric 
              SrRuO$_{3}$/KNbO$_{3}$/SrRuO$_{3}$
              heterostructure. 
              Black curve: planar-averaged $\overline{\rho}_{\rm free}$.
              Red dashed: $\overline{\overline{\rho}}_{\rm free}$,
              nanosmoothed using a Gaussian filter. 
              Blue symbols: finite differences of the local $P_j$ 
              (shown as a black curve in Fig.~\ref{figpol_kno}), 
              calculated using the Wannier-based layer polarization described
              in Sec.~\ref{sec:localpolwan}.}
    \label{fig2}
 \end{figure}

 First we show that the non-vanishing LDOS at the Fermi level results in a 
 sizeable spill-out of conduction charge into the ferroelectric film.
 To that end, we plot $\overline{\rho}_{\rm free} (z)$, which represents the
 planar average of the artificially populated part of 
 the KNbO$_{3}$ conduction band,
 and the corresponding nanosmoothed version, 
 $\overline{\overline{\rho}}_{\rm free} (z)$, in
 Fig.~\ref{fig2} respectively as black continuous and red dashed lines.
 The additional electron density in the ferroelectric region is apparent,
 and reaches a maximum of about 0.15 electrons in the central perovskite 
 unit cell.
 Such a density is significant -- it can be thought as resulting from an
 unrealistically large doping of, e.g. one Sr$^{2+}$ cation every six
 or seven K$^+$ ions.
 However, unlike in a doped perovskite, the spurious electron spill-out
 here is not compensated by an appropriate density of heterovalent cations.
 The system is therefore not locally charge neutral, 
 and as a consequence strong, non-uniform electric fields 
 arise in the insulating film that act on the
 ionic lattice.

 \begin{figure}
    \begin{center}
       \includegraphics[width=0.35\textwidth,clip]{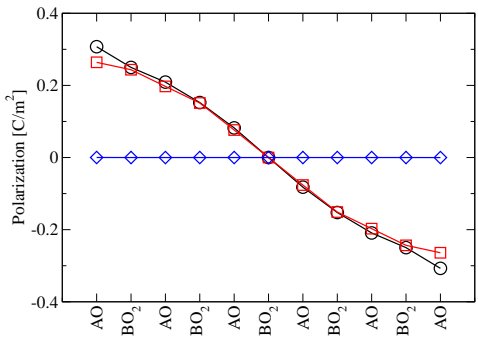}
    \end{center}
    \caption{ (Color online) 
              Local polarization profile in the 
              SrRuO$_{3}$/KNbO$_{3}$/SrRuO$_{3}$ capacitor.
              Black circles: polarization from Wannier-based layer polarizations.
              Red squares: approximate polarization from ``renormalized'' Born
              effective charges (see Sec. \ref{sec:localpolborn}).
              Analogous results
              for a paraelectric SrRuO$_{3}$/BaTiO$_{3}$/SrRuO$_{3}$ 
              capacitor are shown for comparison (blue diamonds).}
       \label{figpol_kno}
 \end{figure}

\begin{figure}
    \begin{center}
        \includegraphics[width=0.35\textwidth,clip]{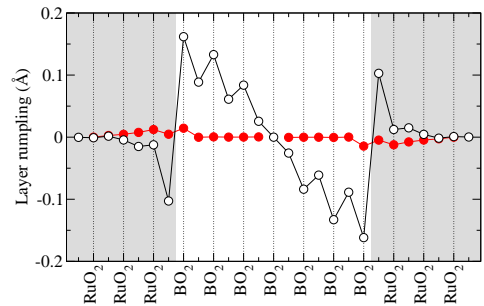} 
    \end{center}
    \caption{ (Color online) Layer rumplings (cation-oxygen vertical
              relaxations) in the centrosymmetric KNbO$_{3}$/SrRuO$_{3}$
              (black line, empty circles) and 
              PbTiO$_{3}$/SrRuO$_{3}$ (red line, filled circles) capacitors.
              Dashed vertical lines indicate the location of the BO$_2$
              planes. The shaded areas correspond to the SrRuO$_3$ electrode
              region.}
    \label{figcoo}
 \end{figure}

 In order to elucidate how the underlying polarizable material
 responds to such an electrostatic perturbation, we plot in
 Fig.~\ref{figpol_kno} the effective polarization profile in the
 KNbO$_{3}$ film calculated in two ways, (i) the rigorous Wannier-function
 analysis of the layer polarizations 
 and (ii) the approximate expression
 based on the renormalized bulk dynamical charges.
 The matching between the curves is excellent, indicating that
 the approximate $Z^*$-based formula provides a reliable estimate
 of $P(z)$; this suggests that the electrostatic screening is indeed 
 dominated by structural relaxations, as 
 anticipated in Sec.~\ref{sec:theory}, and as expected in a 
 ferroelectric material.
 To substantiate this point, we compare in
 Fig.~\ref{figcoo} the relaxed layer rumplings in 
 KNbO$_{3}$/SrRuO$_{3}$ to those of the non-pathological case, 
 PbTiO$_{3}$/SrRuO$_{3}$, discussed in Sec.~\ref{sec:unpolarnonpath}.
 The KNbO$_{3}$ film is characterized by strong non-homogeneous 
 distortions, which are consistent with the polarization pattern
 shown in Fig.~\ref{figpol_kno}. Conversely, the distortions are
 negligible in the PbTiO$_{3}$/SrRuO$_{3}$ capacitor, where all the
 oxide layers are essentially flat. 

 The polarization profile $P_j$ is characterized by a uniform, negative
 slope. This nicely confirms the prediction of our semiclassical 
 analysis in Sec.~\ref{sec:theory} of a uniform linear decrease
 of $D(z)$ throughout the film. $P_j$ varies from 0.3 to -0.3 C/m$^2$ 
 when moving from the bottom to the top interface.
 Note that such spatial variation is completely absent in, e.g., isostructural
 paraelectric BaTiO$_3$/SrRuO$_3$ (diamonds in Fig.~\ref{figpol_kno}), 
 and PbTiO$_3$/SrRuO$_3$ [Fig. \ref{fig:unpolldapto}(c)] capacitors,
 where the profile is remarkably flat with $P$ vanishing throughout the film.
 We stress that the non-uniform perturbation experienced by 
 KNbO$_{3}$/SrRuO$_{3}$ is qualitatively different from a 
 ferroelectric distortion, which involves an almost perfectly 
 rigid displacement of the ionic sublattices: in absence of 
 space-charge effects, a macroscopically uniform rumpling 
 pattern across the film is typically found.~\cite{Stengel-09.2}

 To demonstrate that the spatial variation in $P(z)$ is directly
 related to $\rho_{\rm free}$ according to Eq.~(\ref{eq:maxwell}),
 we perform a numerical differentiation 
 of the polarization profile derived from the Wannier-based layer polarizations.
%
%
 The result, plotted in Fig.~\ref{fig2} as a blue line,
 shows an essentially perfect match between $dP/dz$ and $-\rho_{\rm free}$
 illustrating the fact that the polarization profile
 is really a consequence of KNbO$_{3}$ responding to the spurious
 population of the conduction band, rather than of interface bonding
 effects.~\cite{Duan-06}

\subsubsection{BaTiO$_3$/Pt}
\label{sec:btopt}

\begin{figure}[!t]
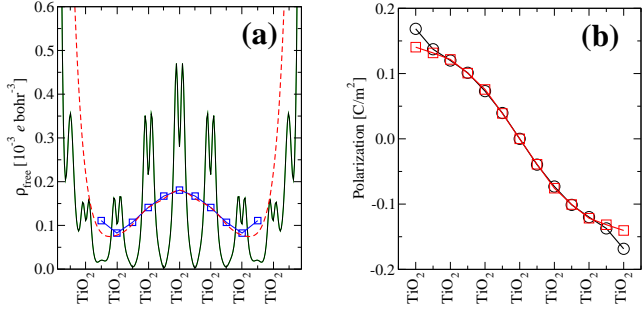

    \begin{center}
       \includegraphics[width=0.22\textwidth,clip]{fig12a.eps} \hspace{0.02\textwidth}
       \includegraphics[width=0.22\textwidth,clip]{fig12b.eps} 
    \end{center}
    \caption{ (Color online) (a) Calculated free charge  
               and (b) local polarization profile
               for a paraelectric 
               Pt/BaTiO$_3$/Pt capacitor with TiO$_2$-type interfaces. 
               All symbols have the same meaning
               as in Fig.~\ref{fig2}  and Fig.~\ref{figpol_kno}.}
    \label{figpol_bto}
 \end{figure}

 We next present results of an analogous investigation for a paraelectric 
 (BaTiO$_3$)$_m$/(Pt)$_n$ capacitor, with $m=8.5$ and $n=11$. We consider 
 symmetric TiO$_2$ terminations, with the interfacial O atoms in the on-top 
 positions.
 (Note that this interface structure is different than the AO-terminated
 films simulated, e.g. in Refs.~\onlinecite{Stengel-09.2} and
 \onlinecite{nature_mat}, where a Schottky-like band offset was found.)
 We find this interface to have a pathological band alignment,
 similar to the KNbO$_{3}$/SrRuO$_{3}$ case discussed above.
 The comparative analysis of the bound-charge polarization profile
 and of the excess conduction charge, shown in Fig.~\ref{figpol_bto},
 again shows excellent agreement between 
 $\bar{\bar{\rho}}_{\rm free}(z)$ and the
 compensating bound charge.
 The effect is analogous to KNbO$_{3}$/SrRuO$_{3}$, with an overall 
 magnitude which is smaller by roughly a factor of two; the polarizations
 at the two extremes of the film reach values of about $\pm$0.15 C/m$^2$.

 The almost perfect similarity in behavior between these two chemically 
 dissimilar systems is further proof that the unusual effects described 
 here and in Ref.~\onlinecite{Duan-06} -- the apparent head-to-head 
 domain wall in the ferroelectric film -- have little to do with the 
 bonding at the interface, but are merely a consequence of the artificial 
 charge spill out, as discussed in Sec.~\ref{sec:theory}.

 Before moving on to the next Section we briefly comment 
 on the physical nature of the conduction charge that spills into the
 ferroelectric film. 
 In particular, it is important to clarify that the charge densities
 plotted in Fig.~\ref{fig2} and Fig.~\ref{figpol_bto}(a) indeed 
 originate from population of the conduction band of the insulator,
 and not from metal-induced gap states (MIGS) as some authors have 
 recently argued.~\cite{Wang-10} 
%
 First, all charge density plots show a \emph{maximum} in the 
 middle of the ferroelectric layer, rather than a minimum, which one
 would expect if the former hypothesis were true, given the evanescent 
 character of the MIGS. 
 Second, if MIGS were present they would be clearly identifiable in the 
 local density of states; however, the LDOS plotted in 
 Fig.~\ref{fig:knoldos} shows no evidence of quantum states lying within 
 the energy gap of the KNbO$_3$ film.
%
 Therefore, we must conclude that these are genuine conduction band states, 
 and not MIGS. The maximum of $\bar{\bar{\rho}}_{\rm free}$ in the middle of the 
 ferroelectric film can be interpreted either as a quantum confinement
 effect [the lowest-energy solution of the electron-in-a-box problem is 
 indeed a sine function with a shape reminiscent of the 
 $\bar{\bar{\rho}}_{\rm free}$ plots of Fig.~\ref{fig2} 
 and Fig.~\ref{figpol_bto}(a)], and/or as a result of the dielectric 
 nonlinearities discussed in Sec.~\ref{sec:bettermodel}.

\subsection{Estimating the ``pre-spill'' band offset}

 \begin{figure}
    \begin{center}
       \includegraphics[width=2.0in]{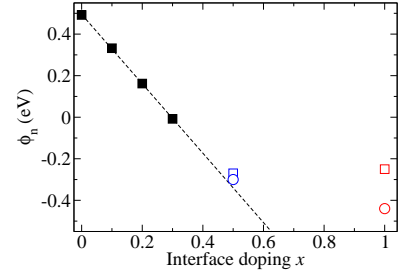}
    \end{center}
    \caption{(Color online) $n$-type Schottky barrier as a function of 
             interface doping in
             KNbO$_3$/AO-terminated SrRuO$_3$, where A is a fictitious atom
             with atomic number $Z = 37 + x$.
             Only the Sr atoms at the interfacial layer are replaced by
             this fictitious atom.
             The dashed line is a linear regression of the 
             data between $x=0$ and $x=0.3$, where the interface
             is non-pathological from the band alignment point of view.
             Blue and red empty symbols represent, respectively, the results
             for $x$ = 0.5 and $x$ = 1.0, where the interface is already
             pathological. All values were obtained from Eq.~(\ref{eq:semicore}),
             using either the Nb(4s) (squares) or the O(2s) (circles) semicore 
             peaks of the central NbO$_2$ layer as a reference.
             \label{fig:extrap}}
 \end{figure}

 We mentioned in Sec.~\ref{sec:theory} that, whenever an 
 electrode/ferroelectric interface enters the pathological spill-out 
 regime, the transfer of charge into the conduction band of the 
 insulator produces an upward shift of the CBM. This effect prevents 
 a direct, unambiguous determination of the interface parameter $\phi_n^0$. 
%
%
%
 To circumvent this problem, and obtain an approximate estimate of the
 negative ``pre-spill'' Schottky barrier $\phi_n^0$, we use an approach 
 inspired by a recent work.~\cite{Burton-10}
 The authors of Ref.~\onlinecite{Burton-10} show that the Schottky 
 barrier at the interface between a perovskite insulator (SrTiO$_3$) and a 
 perovskite electrode (La$_{0.7}$A$_{0.3}$MnO$_{3}$, where
 A is Ca, Sr, or Ba) evolves linearly as a function of the
 compositional charge of the interface layer. (This interface layer
 is of the type La$_x$Sr$_{1-x}$O, where $x$ interpolates between a +3
 and a +2 cation.)
 Of course, this linear behavior refers to a range of $x$ values where the
 interface is non-pathological; our arguments indicate that as soon as 
 the system enters the spill-out regime, the value of $\phi_n$ saturates
 to a nearly constant value.
 Based on this observation, if one knows the linear behavior of $\phi_n$ in
 a range of $x$ values for which the interface is non-pathological, one can
 extrapolate this straight line to the values of $x$ which cannot be 
 directly calculated, and obtain an estimate for $\phi_n^0$.

 \begin{figure}
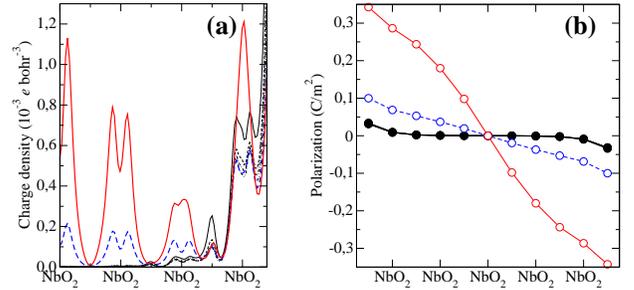

    \begin{center}
       \includegraphics[height=1.5in,clip]{fig14a.eps}  \hspace{0.1in}
       \includegraphics[height=1.5in,clip]{fig14b.eps} \\ 
    \end{center}
    \caption{ (Color online) (a) Conduction electrons, and 
              (b) local (Wannier-based) 
              polarization profiles extracted from the calculations with
              $x=$0.0, 0.1, 0.2 and 0.3 (filled circles, thin black curves); 
              0.5 (empty blue circles, dashed blue curve) 
              and 1.0 (empty red circles, solid red curve).
              In (a) only half of the KNbO$_3$ film is shown. 
              \label{fig:crossover}}
 \end{figure}

 We apply this strategy to the same KNbO$_3$/SrRuO$_3$ capacitor system 
 described in Sec.~\ref{sec:knosro}. To tune the interface charge, we 
 replace the Sr cation in the interface SrO layer with a fictitious atom
 of fractional atomic number $Z=37+x$. $x=1$ corresponds to the 
 example already shown in Sec.~\ref{sec:knosro}, 
 with a charge-neutral SrO interface 
 layer, and $x=0$ corresponds to a RbO layer of net formal charge -1. 
 The results for the Schottky barrier are plotted in Fig.~\ref{fig:extrap}.
 The region from $x=0.0$ to $x = 0.3$ is non-pathological and shows an
 almost perfectly linear evolution of $\phi_n$ (dashed line). 
 By extrapolating this linear trend to $x=1$ we obtain $\phi_n \sim -1.2$ eV,
 which is about 1 eV lower than the value calculated from first principles.
 This confirms the remarkable impact of the space-charge effects described in
 Sec.~\ref{sec:paraspill}.
 Assuming a polarization of $\sim 0.3$ C/m$^2$ for KNbO$_3$ near the interface,
 a potential drop of 1 eV would be accounted for by an
 effective screening length 
 of 0.3 \AA{} at the electrode interface. This value is quite reasonable, 
 and similar in magnitude to those reported in Table~\ref{tab:deltap} .

 In order to examine the crossover between the Schottky (non-pathological)
 and the Ohmic (pathological) regimes in terms of the analysis tools 
 developed in this work, we plot in Fig.~\ref{fig:crossover} the
 polarization profiles and the density of conduction electrons 
 for each of the calculations summarized in Fig.~\ref{fig:extrap}.
 These plots confirm that from $x=0$ to $x=0.3$ 
 the capacitors are non-pathological,
 with absence of conduction charge in the insulating region [panel (a), thinner
 lines] and a flat polarization profile [panel (b), filled circles -- all these 
 curves overlap on this scale]. 
 Conversely, at $x=0.5$ the conduction band starts populating significantly 
 [thicker dashed blue line in (a), empty blue circles in (b), 
 note that the corresponding points in 
 Fig.~\ref{fig:extrap} starts to depart from the linear regime].
 At $x=1.0$ population of the conduction band has become dramatic, and 
 so is the corresponding slope in the polarization profile. The
 departure from linearity in Fig.~\ref{fig:extrap} is correspondingly
 large.
 Note that the use of either the Nb(4s) or the O(2s) semicore peaks 
 in Eq.~(\ref{eq:semicore}) yields identical results in the non-pathological
 regime (the filled squares and circles overlap in Fig.~\ref{fig:extrap}).
 Conversely, the result \emph{depends} significantly on this (completely 
 arbitrary) choice at $x=0.5$, and even more so at $x=1.0$ (the circles
 and squares split).
 This is another proof that in the pathological regime the band lineup 
 is ill-defined -- due to the electrostatic effects discussed throughout
 this work, the LDOS does not converge to a bulk-like value in
 the center of the KNbO$_3$ film (see Fig.~\ref{fig:knoldos}),
 and there is no obvious reference energy to determine the offset.

\section{Results: Ferroelectric capacitors}
\label{sec:results-ferro}

 As discussed in the Introduction, although some of the unpolarized
 reference structures (e.g. the PbTiO$_3$/SrRuO$_3$ interface) appear
 artifact-free within LDA, because of the strong dependence of the 
 Schottky barrier on $D$ [Eq.~\ref{eq:phid}] they might become problematic when the 
 constraint of mirror symmetry is lifted and the system is polarized.
 To address this issue, in this section we first use the
 fixed-$D$ strategy described in Sec.~\ref{sec:constrainedd} to
 explore the behavior of the PbTiO$_{3}$/metal interface over a wide range
 of polarization states.
 Then, we will demonstrate that the behavior that we calculate using
 the fixed-$D$ method
 corresponds exactly to that of a true short-circuited
 capacitor by performing more ``standard'' large-scale calculations
 for a few selected thickness values.

\subsection{Open-circuit calculations}
\label{sec:opencircuit}

 \begin{figure}
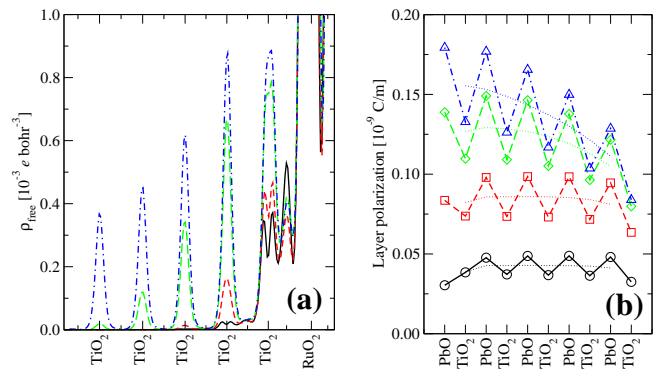

    \begin{center}
       \includegraphics[height=1.9in,clip]{fig15a.eps} 
       \hspace{5pt}
       \includegraphics[height=1.9in,clip]{fig15b.eps} \\
    \end{center}
    \caption{\label{figptoall} (Color online) Results for the
             polarized PbTiO$_3$/SrRuO$_3$ interface for increasing
             polarization of the film.
             (a) planar averaged $\overline{\rho}_{\rm free}$. 
             Black, red, green and blue curves
             correspond to the results for $d=$0.20, 0.40, 0.60 and 0.74 $e$,
             respectively. The sharp peaks in $\overline{\rho}_{\rm free}$ 
             correspond to the
             Ti ions in the PbTiO$_3$ film.
             (b) layer polarizations from the Wannier-based analysis.
             Same color code as in (a).}
 \end{figure}

 \begin{figure}
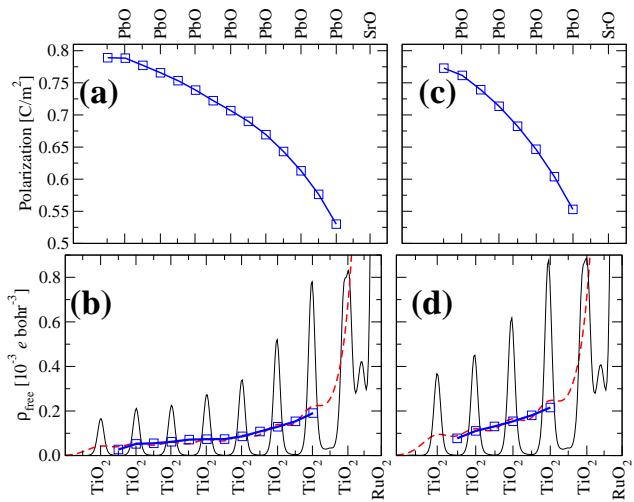

    \begin{center}
       \includegraphics[height=1.27in,clip]{fig16a.eps} \\
       \vspace{0pt}
       \includegraphics[height=1.3in,clip]{fig16b.eps} \\
    \end{center}
    \caption{\label{figptolp} (Color online) Calculated results for the
             fully polarized PbTiO$_3$/SrRuO$_3$ interface at $d=0.74$.
             (a) local polarization
             from the Wannier-based layer polarizations, 
             and (b) planar averaged $\overline{\rho}_{\rm free}$ 
             (black curve),
             macroscopically averaged $\overline{\overline{\rho}}_{\rm free}$ 
             (red dashed curve), 
             and finite differences
             of the polarization shown in the panel (a) (blue squares)
             for an $m$ = 8-unit
             cell thick PbTiO$_{3}$ film. 
             Panels (c) and (d) are the corresponding figures for 
             an $m$ = 5-unit cell thick PbTiO$_{3}$ film.
             The sharp peaks in $\overline{\rho}_{\rm free}$ correspond to the 
             Ti ions in the PbTiO$_3$ film.}
 \end{figure}

 We construct a vacuum/PbTiO$_3$/SrRuO$_3$ heterostructure as explained
 in Sec.~\ref{sec:constrainedd}.
 The reduced macroscopic displacement field~\cite{fixedd}, $d=DS$, is controlled 
 by substituting the Ti at the PbTiO$_{3}$/vacuum interface with a 
 fictitious cation of atomic number $Z_{left}=40+d$ (i.e. Zr for $d$=0).
%
%
 The thickness of the PbTiO$_{3}$ slab is set to 5 unit cells, 
 and that of SrRuO$_{3}$ to 4; other computational parameters 
 are kept the same as in the rest of this work.
 We considered four different values of $d$: 0.2, 0.4, 0.6 and
 0.74, the latter one corresponding to the ferroelectric ground state
 of PbTiO$_3$ at the SrTiO$_{3}$ in-plane lattice constant.
 In each case, we verify by examining the LDOS that the free
 surface remains locally insulating; therefore, the macroscopic
 $D=d/S$ in the film corresponds exactly to the value enforced by
 the artificial pseudopotential.

 The evolution of $\rho_{\rm free}$ and of the Wannier-based layer
 polarization profiles for $0.2\le d\le 0.74$ is shown in Fig.~\ref{figptoall}.
 It is apparent from the plots of $\rho_{\rm free}$ that already for the
 smallest value of the polarization [$d=0.20$, black curve in 
 Fig.~\ref{figptoall}(a)]
 the TiO$_2$ layer closest to the electrode has an important density of
 conduction electrons.
 This is expected, as the evanescent tails of the metal-induced gap states
 (MIGS) penetrate into the insulating region for some distance at any
 metal/insulator junction.
 However, these states do not propagate very far, and already at the
 second TiO$_2$ layer they are barely noticeable on the scale of the
 plot.
 At $d=0.4$ [red curve in  Fig.~\ref{figptoall}(a)] 
 the peak on the second TiO$_2$ layer
 significantly increases in magnitude, and a new small peak appears
 at the third TiO$_2$ layer.
 Analysis of the local density of states (not shown) shows that
 these new peaks are conduction band states of PbTiO$_3$, rather than
 evanescent SrRuO$_3$ states. The reason why $\rho_{\rm free}$ decays
 relatively fast when moving into the insulator is due here to the internal 
 field in PbTiO$_{3}$, which generates a confining wedge-like potential.
 We stress again that this mechanism is fundamentally different from the 
 usual quantum-mechanical damping of the MIGS that fall in a forbidden 
 energy window of the insulator.
 We identify this mechanism with the onset of the Schottky breakdown, which
 becomes increasingly apparent if the polarization of the film is further
 increased to $d=0.60$ [green curve in  Fig.~\ref{figptoall}(a).] 
 As in the discussion of the paraelectric capacitors, the presence of the
 space charge is reflected in the progressive ``bending'' of the layer
 polarization profile [Fig.~\ref{figptoall}(b)].

 At $d=0.74$, the population of the conduction band becomes rather dramatic,
 and the charge distributes over the whole film.
 Here, the space charge is no longer confined by the depolarizing field:
 in the fully polarized ferroelectric state the internal field of PbTiO$_{3}$ 
 vanishes.
 Therefore, the intrinsic carriers are only loosely bound to the interface
 by the band bending effect, analogous to the mechanism that confines
 the compensating carriers at the LaAlO$_3$/SrTiO$_3$ interface.~\cite{laosto}
 Since the dielectric permittivity of PbTiO$_{3}$ is rather large, 
 the band bending is very efficiently screened, and the distribution 
 of charge can reach quite far into the insulator.
 To demonstrate this fact, we have repeated the simulation with the same value
 of $d$=0.74, but with a thicker PbTiO$_{3}$ film of 8 unit cells 
 [Fig.~\ref{figptolp} (b)];
 indeed, the conduction electrons redistribute over the whole volume of the
 film to minimize their kinetic energy.
%
 Thus, in close analogy to the LaAlO$_{3}$/SrTiO$_{3}$ case,
 the metallization of the fully polarized PbTiO$_{3}$ film at
 $d$=0.74 can be thought as a form of ``electrostatic doping'' induced
 by spill-out of electrons from the electrode to the 
 PbTiO$_{3}$ conduction band.
 We shall further elaborate on this point in Sec.~\ref{sec:laosto}.

 Fig.~\ref{figptoall}(b) illustrates a further important consequence of 
 the charge spill-out regime, which was mentioned 
 already in Sec.~\ref{sec:ferrospill}: in the pathological regime the 
 dipoles that lie closest to the electrode interface may appear ``pinned'' 
 to a fixed value.
 This is indeed the case for the TiO$_2$ layer adjacent to the electrode,
 which seems to saturate at $\sim 0.08$ nC/m for increasing values of $D$.
 Again, we caution against interpreting this dipole pinning effect as 
 a robust physical result.

\subsection{Short-circuit calculations}
\label{sec:shortcircuit}

 To demonstrate in practice that the conclusions of 
 Sec.~\ref{sec:opencircuit}, inferred by using open-circuit 
 boundary conditions, are directly relevant to short-circuited
 capacitors, we have performed simulations on 
 SrRuO$_{3}$/PbTiO$_{3}$ heterostructures, with 
 $m$ = 12.5 and $n$ = 5.5.
 A soft-mode distortion of the bulk tetragonal
 phase, inducing a polarization perpendicular to the interface, 
 is superimposed on the PbTiO$_3$ layers of the previous
 unpolarized configurations discussed in Sec.~\ref{sec:unpolarnonpath}.
 Then the atomic positions of all the ions,
 both in the ferroelectric and in the metallic electrodes,
 and the out-of-plane stress are relaxed again with the
 same convergence criteria as before.

 \begin{figure}
    \begin{center}
    \includegraphics[width=0.4\textwidth,clip] {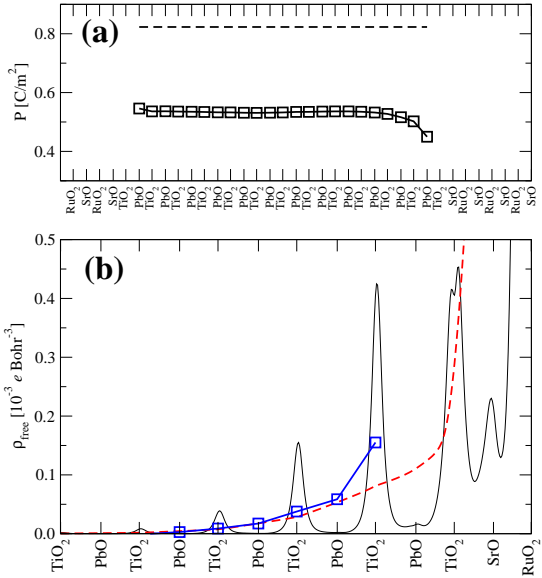}
       \caption{ (Color online) 
                 (a) Profile of the layer-by-layer polarization $\tilde P^Z_j$,
                 defined in Eq. (\ref{eqzstar}), in the relaxed
                 polar configuration of a short-circuited 
                 SrRuO$_{3}$/PbTiO$_{3}$/SrRuO$_{3}$ capacitor.
                 The dashed line represents the bulk spontaneous polarization
                 under the same in-plane strain as in the capacitor.
                 (b) $\bar{\rho}_{\rm free}(z)$ as defined in 
                  Eq. (\ref{eq:planaverhofree}) (black solid line),
                 and its nanosmoothed average $\bar{\bar{\rho}}_{\rm free}(z)$
                 (red dashed line).
                 The blue line represents the profile of the bound charge,
                 computed as a finite-difference derivative of $\tilde P^Z_j$.
         }
       \label{fig:polldapto}
    \end{center}
 \end{figure}

 By means of the approximate Eq. (\ref{eqzstar}), derived in
 Sec. \ref{sec:localpolborn}, we computed the local layer-by-layer
 polarization, $\tilde P^Z_j$, plotted in Fig. \ref{fig:polldapto} (a).
 Far enough from the interface,
 the polarization profile is rather uniform,
 with a polarization that amounts to 0.53 C/m$^{2}$ in PbTiO$_{3}$
 (64 \% of the strained bulk polarization),
 which we identify as the macroscopic $P$ of the PbTiO$_3$ film. 
%
 This corresponds to $d \sim 0.5$, i.e. a value that in
 the open-circuit study of the previous section (see Fig.~\ref{figptoall})
 we found to be already pathological.
 To verify that the same
 happens here, we analyze the density of conduction electrons.
 The planar average of $\rho_{\rm free} (\bf{r})$
 for the relaxed polar configuration is plotted in
 Fig. \ref{fig:polldapto}(b).
 The existence of a charge populating the Ti $3d$ orbitals is evident
 from the peaks of $\bar{\rho}_{\rm free}(z)$ at the TiO$_{2}$ layers, which
 are detectable up to four unit cells away from the interface.
 Indeed the profile of the conduction charge appears to be intermediate
 between the $d=0.4$ and $d=0.6$ cases of Sec.~\ref{sec:opencircuit},
 consistent with the present estimate $d\sim 0.5$.
%

 As we already anticipated in the previous Sections, 
 $\rho_{\rm free}$ is responsible for non-trivial lattice
 relaxations, which act to screen the electrostatic
 perturbation.
 Fig. \ref{fig:polldapto}(a) indeed shows
 a small bending of the local polarization profile,
 starting roughly three unit cells away from the
 top interface and with a negative slope of the local polarization, 
 $\tilde{P}_{j}^{Z}$. 
%
%
%
 To prove that such a spatial variation of $P(z)$ 
 [which in PbTiO$_3$ provides a reasonably accurate estimate of the local 
 electric displacement $D(z)$] is a direct consequence of the presence of
 the non-vanishing conduction charge [recall Eq.~(\ref{eq:maxwell})], we 
 numerically differentiatiate the polarization profile and compare it 
 with $\bar{\bar{\rho}}_{\rm free}(z)$ in Fig. \ref{fig:polldapto}(b).
 As in the SrRuO$_{3}$/KNbO$_{3}$/SrRuO$_{3}$ unpolarized case
 (see Fig. \ref{fig2}), the bound charge (divergence of
 ${\rm P}$) accurately neutralizes the conduction charge (nanosmoothed 
 profile of $\rho_{\rm free}$).

 \begin{figure}
    \begin{center}
    \includegraphics[width=0.45\textwidth,clip] {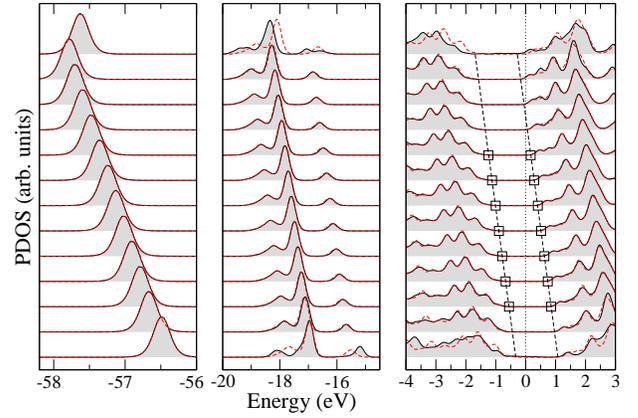} 
       \caption{ (Color online) Layer-by-layer PDOS on the TiO$_2$ layers of the
                 polar SrRuO$_{3}$/PbTiO$_{3}$/SrRuO$_{3}$ 
                 ferroelectric capacitor.
                 Meaning of the lines as in Fig.~\ref{fig:paraldos},
                 but now the PDOS on all the TiO$_{2}$ layers are plotted
                 (there is no longer a mirror symmetry plane).
                 The squares represent the position of the local band edges, 
                 computed following the recipe of Sec.~\ref{sec:best}.
                 The dashed lines are a linear interpolation of the calculated
                 local band edges.
         }
       \label{fig:ferroldos}
    \end{center}
 \end{figure}

 In order to further prove that the present case fits into the model
 description of Sec.~\ref{sec:ferrospill}, in Fig.~\ref{fig:ferroldos}
 we plot the layer-resolved PDOS. The curves were constructed exactly as in 
 Fig.~\ref{fig:paraldos}, except that (i) the capacitor is now polarized; and
 (ii) consistent with the discussion of Sec.~\ref{sec:best} we set up the
 bulk reference calculation by using the PbTiO$_3$ structure extracted from 
 the polarized supercell calculation (i.e. with atomic distortions and 
 out-of-plane strain consistent with a polarization of 0.53 C/m$^2$).
 The agreement is again very good, showing that our approximation of
 neglecting the macroscopic depolarizing field in the bulk reference
 calculation is a reasonable one, and that the most important effects on 
 the LDOS originate from the lattice distortions.
%
 In the capacitor we clearly distinguish two regions. In the lower part of
 the PbTiO$_3$ film, the PDOS at the Fermi level vanishes, which implies
 that the system is locally insulating. Furthermore, the PDOS in each 
 layer appears rigidly shifted with respect to the neighboring two layers, 
 consistent with the presence of a depolarizing field. In the upper 
 region, close to the top electrode, the PDOS crosses the Fermi level 
 and the system is locally metallic. All these features are in full 
 agreement with the scheme drawn in Fig.~\ref{fig:ferrospill}.

 In Fig.~\ref{fig:ferroldos} we also plot the estimated band edges for
 each layer, $E_{\rm V,C}(j)$, obtained from Eq.~(\ref{eq:semicore}).
 We used the semicore Ti(3s) peak at each layer as $E_{\rm sc}(j)$, and
 we calculated the bulk contributions in Eq.~(\ref{eq:semicore}) from
 a non-self-consistent bulk calculation (based on the ground state
 charge density of the bulk reference calculation at $P=$0.53 C/m$^2$
 described above) that included the high-symmetry $k$-points. 
%
 The resulting data points lie very accurately on a straight line. By extrapolating
 this straight line, we see that it crosses the Fermi level near the fourth 
 PbTiO$_3$ cell from the top electrode interface. This illustrates the pathological
 character of the band alignment in this system, consistent with the model of
 Fig.~\ref{fig:ferrospill}.

 As a final remark, we mention that we performed similar calculations
 for a polarized BaTiO$_3$/SrRuO$_3$ capacitor, and found a very similar scenario,
 with the conduction band locally crossing the Fermi level as the mirror symmetry plane 
 is lifted and a spontaneous polarization is allowed to develop.
 In general, the onset of such a pathological regime has important consequences on 
 many physical properties of the capacitor, as we shall discuss in the following
 Section.

\section{Discussion}
\label{sec:discussion}

 In this Section we discuss the important aspects of our work in
 the context of the existing literature. The discussion is organized in
 several categories, corresponding to the different properties 
 of a ferroelectric/electrode interface (or, more generally, of a perovskite 
 material) that might be affected by the (more or less spurious) presence 
 of free charges in the system. 

\subsection{Structural properties of the film}

 The authors of Ref.~\onlinecite{Duan-06} studied KNbO$_{3}$ thin films placed
 between symmetric metallic electrodes (either SrRuO$_{3}$ or Pt) under 
 short-circuit electrical boundary conditions.
 In the SrRuO$_{3}$ case, the layer-by-layer polarization pointed in opposite
 directions at the top and bottom interfaces for all thicknesses,
 creating 180$^\circ$ head to head domain walls,
 which were denominated {\it interface domain walls} (IDW).
 The physical origin of the IDW was attributed to a strong bonding
 between interfacial Nb and O atoms, which would induce a ``pinning''
 of the interface dipoles to opposite values at the top and bottom electrode
 interfaces.

 Here we have demonstrated with analytical derivations and practical examples
 that both the inhomogeneous polarization and the 
 ``dipole pinning'' effect are clear signatures of a pathological 
 band alignment.
 In particular, in an unpolarized KNbO$_3$/SrRuO$_3$ capacitor analogous 
 to those simulated by Duan {\em et al.}~\cite{Duan-06},
 we obtain a monotonously decreasing polarization profile, from ($\sim$0.3 C/m$^2$) at the bottom
 interface to an opposite value of $\sim$-0.3 C/m$^2$ at the top,
 in excellent agreement with the results of Duan and coworkers.\cite{Duan-06}
 In contrast with the conclusion of Ref.~\onlinecite{Duan-06}, however, 
 here we find that the microscopic origin of this strong
 inhomogeneous polarization 
 is the  spillage of charge from the metallic electrode to the bottom
 of the conduction band of KNbO$_{3}$, rather than a bonding effect.

 These findings have important consequences concerning the physical
 understanding of the system with regard to the relevant observables.
 First, the ferroelectric material becomes in fact a metal, and such a device 
 would respond Ohmically with a large direct DC current
 that would make switching difficult or impossible. This questions the 
 appropriateness of interpreting the ``average'' polarization of the film 
 as a macroscopic physical quantity that can be measured in an experiment (see next Section).
 Second, our arguments indicate that two essential factors
 governing the equilibrium free charge distribution (and hence the spatial 
 variation of $P$) are the conduction band structure (e.g. the density of
 states) of the ferroelectric material, and the interface band offset. Both 
 ingredients are absent in
 traditional Landau-Ginzburg models, e.g. those used in Ref.~\onlinecite{Duan-06} 
 to interpret the above data on KNbO$_3$/SrRuO$_3$ capacitors, or in 
 Ref.~\onlinecite{Wang-10} to interpret qualitatively similar results for a 
 hole-doped BaTiO$_3$/SrRuO$_3$ interface. 
 We therefore caution against overinterpreting the results of such models,
 as they might fail at capturing the relevant physics of the free-charge 
 equilibration. 
 A promising route towards overcoming these limitations appears to be the 
 model Hamiltonian approach proposed in Ref.~\onlinecite{laosto}.
 Extending that strategy to the case of a metal/ferroelectric interface will 
 be an interesting subject of further research.

\subsection{Stability of the ferroelectric state}

\begin{figure}
    \begin{center}
    \includegraphics[width=2.5in,clip] {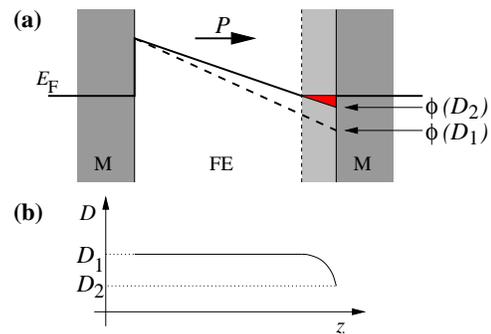}
       \caption{(Color online) Schematic representation of the impact of the
        charge spillage on the ferroelectric stability. M are the metal 
        electrodes, and FE is the ferroelectric film. The polarization
        points to the right.
       \label{fig:stability} }
    \end{center}
 \end{figure}

 The pathological spill-out of charge has important consequences on
 the spontaneous polarization of a ferroelectric capacitor. 
 To give a qualitative flavor of such an effect, we consider the
 case of a capacitor that is only partially metallic, i.e. there is
 a depolarizing field that keeps the carriers confined to the pathological
 side as sketched in Fig.~\ref{fig:stability}(a). 
 We further consider two symmetric electrodes, i.e. characterized by 
 identical values of $\phi_n^0$ (that we assume positive) and $\lambda_{\rm eff}$.
 Assuming a monodomain state, there are then two stable configurations,
 related by a mirror symmetry operation. 
 As $\phi_n^0$ is positive, upon application of an electric field there
 will always be an insulating region in the middle of the film, i.e. 
 the polarization can be switched without passing through a \emph{globally}
 metallic state.

 To appreciate the impact of the charge spill-out on the spontaneous polarization 
 of the film, it is useful to look at the schematic band diagram of 
 Fig.~\ref{fig:stability}(a), where the conduction band bottom goes below 
 the Fermi level in proximity of the right electrode (red area).
 This induces metallicity in a significant portion of the film (light grey 
 shaded area, up to the dashed line).
 Based on our arguments of Sec.~\ref{sec:theory}, the charge spill-out
 is associated with a spatially decreasing $D(z)$ [Fig.~\ref{fig:stability}(b)].
 This, in turn, modifies the interface potential barrier by producing a strong 
 upward shift in energy of the conduction band edge from what one would have if
 $D(z)$ were uniform and equal to the ``physical'' value $D_1$.
 This implies that the charge spill out generally \emph{reduces} the depolarizing 
 field [the ``pre-spill'' estimate is sketched as a thick dashed line in 
 Fig.~\ref{fig:stability}(a)], and hence \emph{overstabilizes} the ferroelectric state.
 This is what one intuitively expects -- population of the conduction band 
 constitutes an additional channel for screening the polarization charge, and
 this cooperates with the metallic carriers of the electrode.
 This, however, contrasts with the conclusions of Ref.~\onlinecite{Wang-10}, where
 it was argued that charge leakage suppresses $P$ by producing a ferroelectrically
 ``dead'' layer.
 These conclusions are based on the assumption that the physically measurable 
 $P$ is the \emph{average} polarization, $\langle P \rangle$, taken over
 the whole volume of the film.
 As the polarization is locally reduced near a pathological interface, 
 charge spill-out indeed results in a reduced $\langle P \rangle$.

 Is it justified, though, to assume that $\langle P \rangle$ is the physically
 relevant quantity in the capacitor? Does $\langle P \rangle$, in other words,
 reflect what is experimentally measured? 
 In an experiment one measures the time integral of the transient current density, 
 $\Delta j$, that flows through the capacitor during the switching process. 
 $\Delta j$ does \emph{not} relate to $\langle P \rangle$. Under the hypothesis
 that at least a portion of the film remains insulating throughout switching, it 
 rigorously follows from the modern theory of polarization~\cite{ferro:2007} that 
 $\Delta j=\Delta D=2|D|$; $D$ is the  value of the (locally uniform)
 electric displacement deep in the insulating region, indicated as $D_2$ in 
 Fig.~\ref{fig:stability}.
 (We assume for simplicity that $D=0$ in the paraelectric reference state.)
 Therefore, observing that $\langle P \rangle$ is reduced upon charge leakage
 does not reflect the true \emph{physical} effect of the pathological band
 alignment, which is an artificial enhancement of the spontaneous $P$ via the 
 reduction of the depolarizing field illustrated above.

 A large number of works~\cite{Sai/Rappe:2005,Elsasser:2006,Al-Saidi-10,Umeno-09}
 have investigated the stability of PbTiO$_3$-based capacitors, and it is 
 impossible here to discuss in detail whether and how the above band-alignment
 issues might have affected each of them (for instance, regarding the polarization 
 enhancements reported in Ref.~\onlinecite{Sai/Rappe:2005}). 
 We limit ourselves to observe that, due to the large spontaneous polarization
 of PbTiO$_3$, the possible consequences of having a pathological ferroelectric 
 state need to be taken seriously into account in the analysis, as we showed
 for the example of SrRuO$_3$ electrodes in Sec.~\ref{sec:results-ferro}.

 \subsection{Transport properties in the tunneling regime}

 Ferroelectric capacitors have been explored as potential tunneling 
 electroresistance devices,\cite{Zhuravlev_et_al:2005} and many recent 
 calculations focused on the calculation of the conductance by means of 
 first-principles methods.
 Metallicity and spill-out of electrons is a serious potential issue
 in this context, as the calculated conductance can potentially be affected
 by the presence of space charge in the system, in a way which is difficult
 to predict. 
 The recent work of Velev \emph{et al.} \cite{Velev-07} appears to be
 concerned by these issues, as it focuses on TiO$_2$-terminated Pt/BaTiO$_{3}$/Pt 
 capacitors. Indeed, we find (see Sec.~\ref{sec:btopt}) 
 that this interface is problematic already in the centrosymmetric paraelectric case. 
 While we have not explored the ferroelectric regime in this system, based on 
 the imperfect screening arguments of Sec.~\ref{sec:theory} (the lineup depends 
 linearly on $P$ around the paraelectric reference phase) we expect the spill-out 
 effect to become worse at least at one of the two interfaces when the capacitor 
 is polarized.

 In fact, the metallicity of the ferroelectric film seems to be confirmed by 
 the data presented by the authors: In Fig. 2(a-b) of Ref.~\onlinecite{Velev-07} 
 the conduction band minimum (CBM) of the central BaTiO$_3$ cell appears to be 
 degenerate or lower than the Fermi level, and in Fig. 1 of the same paper
 the atomic displacements of the ferroelectric phase seems to be 
 strongly asymmetric, consistent with our speculations. 
 While we cannot draw a definitive conclusion (our computational 
 setup slightly differs from that of Ref.~\onlinecite{Velev-07}), 
 our analysis highlights the crucial importance of the band alignment issue,
 and the necessity of performing an adequate and convincing 
 assessment of its impact on the results (e.g. the conductance) 
 in each case.

%
%
%
%

\subsection{Interface magnetoelectric effects}

 Magnetoelectricity is one of the emerging topics in oxide research.
 Despite the intense efforts, one of the main limiting factors still 
 persists: bulk materials displaying a robust magnetoelectric effect are 
 notoriously difficult to find. To work around this problem, several 
 researchers have been looking for alternative solutions by exploring 
 heterostructures and composite materials.
 An interface, due to its lower symmetry, might allow for physical 
 response properties that are absent in the parent compounds.
 A promising route to interfacial ME coupling that has been proposed 
 recently~\cite{natnano_2008} is mediated by charge. The polarization
 of the ferroelectric (or dielectric) lattice produces a bound charge at
 the interface, that is screened by the carriers of the metal. If these
 carriers are spin-polarized, as in a ferromagnet, there will be a net
 change in the magnetization. 

 It is easy to see that the band-alignment issues that we discuss in 
 this work have direct and important implications for the calculation of 
 the carrier-mediated interface ME coefficient.
 In the pathological regime, the
 calculated (magnetic) response will most likely be suppressed, as the
 spill-out charge, rather than the spin-polarized carriers in the electrode,
 will screen the applied bias potential (or the ferroelectric polarization).
 This speculation is directly relevant for interpreting the results of 
 Yamauchi \emph{et al.}\cite{Yamauchi-07} on BaTiO$_3$ films
 sandwiched between Co$_{2}$MnSi (Heusler alloy) electrodes.
%
 Depending on the termination, two qualitatively different
 behaviors were reported: the MnSi/TiO$_{2}$ interface results
 in a pathological band alignment and a strongly
 non-homogeneous local polarization profile; conversely, neither is
 present in the capacitor with the other type of termination,
 which has symmetric Co/TiO$_2$ interfaces.
 A very small magnetoelectric response was reported for the
 MnSi/TiO$_{2}$ case (contrary to the Co/TiO$_2$ case), in qualitative 
 agreement with our arguments above.

 Other recent studies,\cite{Fechner-08,Fechner-10} focusing on ME effects 
 in thin Fe film deposited on ATiO$_3$ (A=Ba,Pb,Sr), also reported strongly 
 non-uniform polarization profiles in the ferroelectric film (e.g. Fig. 3 of 
 Ref.~\onlinecite{Fechner-10}).
 This suggests that also the ATiO$_3$/Fe interface might be concerned by
 the band-alignment issues discussed in this work, with potential impact
 on the physical observables. Our analysis tools should help clarify these
 issues in the above systems and in the Fe/BaTiO$_3$/Fe capacitors of 
 Ref.~\onlinecite{Duan-06.2}.

\subsection{Schottky barriers}

 Direct calculations of Schottky barriers at metal/ferroelectric interfaces
 are, among the many useful physical properties of these junctions, those that
 are most directly affected by the issues we discuss here.
 The consequence of a pathological band alignment is that the estimated
 Schottky barrier is no longer a physically meaningful interface property,
 but is influenced by macroscopic space-charge phenomena.

 A rather comprehensive work on the SrTiO$_3$/transition metal interface was 
 recently reported in Ref.~\onlinecite{Mrovec-09}.
 Without going into too detailed an analysis of the results, we limit 
 ourselves to noting that many of the reported $p$-type 
 SBH for TiO$_2$- or SrO-terminated interfaces are very close to, or
 sometimes well in excess of 1.8 eV. 
 Considering that the LDA/GGA fundamental gap
 of SrTiO$_3$ is around 1.8 eV, the actual $n$-type SBH of the 
 calculation (i.e. not the value corrected with the experimental
 band gap) is close to zero or negative. Therefore, charge spill 
 out is a concrete and likely possibility for many of the 
 investigated structures.

 Note that, contrary to the case of oxide electrodes, ideal interfaces between
 SrTiO$_3$ and simple metals tend to have a smaller 
 $\lambda_{\rm eff}$.~\cite{nature_2006}
 This implies that the effects of the electrostatic re-equilibration described 
 in Sec.~\ref{sec:theory} might be somewhat less dramatic, and the values
 of the self-consistent $\phi_n$ closer to $\phi_n^0$.
 This suggests that the trends and the conclusions reported
 in Ref.~\onlinecite{Mrovec-09} are likely to be robust with
 respect to the issues described in this work. 
 However, a more detailed analysis would be certainly interesting
 in order to assess their impact at the quantitative level.

\subsection{Relationship to LaAlO$_{3}$/SrTiO$_{3}$}
\label{sec:laosto}

 Many of the analysis tools developed in this work are not limited to ferroelectric 
 capacitors, but can be readily extended to other systems where free-charge doping 
 of a band insulator plays a central role.
 An excellent example, where the interpretation of the observed effects is still
 widely debated, is two dimensional 
 conducting electron gas (2DEG) that forms at the polar LaAlO$_3$/SrTiO$_3$ 
 interface.~\cite{Ohtomo-04}
 A central problem is the determimation of the physical effects 
 governing the confinement and equilibrium distribution of the 2DEG.
 Some authors\cite{Janicka-09} propose a mechanism for the confinement of the gas
 based on the formation of metal induced gap states (MIGS) in the band gap 
 of SrTiO$_{3}$.
 Other authors,\cite{Copie-09} however, explain the experimental observations in
 terms of a semiclassical Thomas-Fermi model that is analogous 
 to that described in Sec.~\ref{sec:theory}, and where the MIGS are 
 completely absent.
 Answering the question of whether the MIGS play an important role in this
 system involves a careful analysis of the local electronic properties, and
 more specifically of the LDOS.~\cite{Janicka-09} In this sense, 
 the methodology discussed in Sec.~\ref{sec:best} appears ideally
 suited to clarifying this issue.

 \begin{figure} 
    \begin{center}
    \includegraphics[width=3.0in,clip] {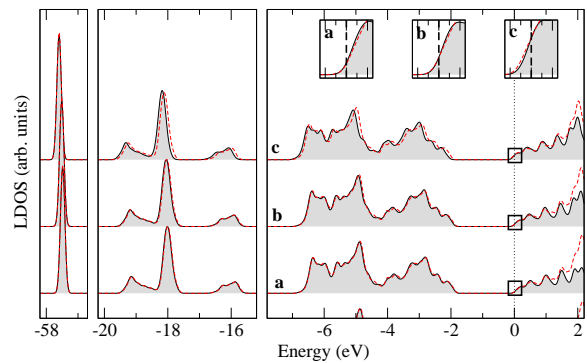}
       \caption{ LDOS of selected TiO$_2$ layers in the electrostatically
               doped LaAlO$_3$/SrTiO$_3$ system. The insets are a blow up of 
               the regions indicated by the small squares. The energy scale
               of the insets is comprised between -0.25 eV and 0.25 eV.               
       \label{fig:laosto} }
    \end{center}
 \end{figure}

We base our analysis on the calculations done in Ref.~\onlinecite{laosto}, 
with a 24-cell thick SrTiO$_3$ slab and a 3-cell LaAlO$_3$ overlayer. 
(This calculation was performed with a $12\times 12$ Monkhorst-Pack sampling
of the surface Brillouin zone, and with a Gaussian smearing of 0.1 eV; 
full details of the computational parameters are reported in Ref.~\onlinecite{laosto}.)
The boundary conditions are set to $D_{\rm STO}$=0, $D_{\rm LAO}=-e/2S$,
and are equivalent to those of the symmetric superlattice used by 
Janicka {\em et al.}~\cite{Janicka-09}
In Fig.~\ref{fig:laosto} we show the LDOS corresponding to the TiO$_2$ layers 
number 15 (curve a), 10 (b) and 5 (c), where layer 1 is adjacent to the LAO
interface.
On top of each curve we superimpose the bulk TiO$_2$ LDOS, that we align with
the supercell LDOS by matching the semicore Ti$(3s)$ peak at $\sim-57.5$ eV.
(As in earlier works the O$(2s)$ peak was used as a reference, we also show the
O$(2s)$-derived feature, which is located at about -18 eV.)
The matching is excellent in all cases, especially in the layers lying furthest
from the interface where the effect of the structural distortions and free charge
are less pronounced.
(Note that we performed the bulk calculation with a $k$-point mesh that accurately
matches the one used in the supercell calculation. Also, in the construction
of the LDOS curves we used the same Gaussian smearing function of width 0.1 eV,
corresponding to the smearing used to relax the self-consistent ground state of 
the supercell structure.)
In the insets we show a blow-up of the conduction band edge, which goes below the
Fermi energy in agreement with the semiclassical arguments of Ref.~\onlinecite{Copie-09}
and of our Sec.~\ref{sec:theory}.
Clearly, our plots do not show any evidence for MIGS in the energy gap,  
contrary to the conclusions of Janicka {\em et al.}~\cite{Janicka-09}

To reconcile this discrepancy, we can speculate that the LDOS curves 
presented in Ref.~\onlinecite{Janicka-09} might have been constructed with 
a substantially larger smearing width than ours, and this might have 
precluded an accurate identification of the band edges.
We believe that the technique presented here (of superimposing an appropriately 
constructed bulk LDOS on top of the supercell curves) provides a very practical
means of minimizing systematic errors in the analysis of the results.

\section{Conclusions}
\label{sec:conclusions}

 Due to its accuracy and efficiency, density functional theory has emerged
 as the method of choice for studying ferroelectric oxides from first-principles.
 This predominance has been reinforced since the early 1990s
 by the many successes achieved in the determination of the structural, 
 energetic, piezoelectric, and dielectric properties at the bulk level.
 In the last few years, those efforts have evolved to address
 the behaviour of the functional properties in
 thin films and superlattices, including in same cases
 (for instance, in the study of ferroelectric capacitors)
 the presence of metal/insulator interfaces.

 For a reliable prediction of the functional properties of
 these devices, the atomic displacements, distortions of the unit cell,
 the electronic structure and the band gap
 have to be accurately described simultaneously.
 However, the proper DFT treatment of such interfaces is complicated by 
 the so-called ``band-gap problem'',
 which might produce a pathological alignment between the Fermi level of the 
 metal and the conduction band of the insulator, thus precluding
 explicit DFT investigation of many systems of practical interest.
 In this work we provide useful guidelines to identify such a pathological 
 scenario in a calculation by examining its main physical consequences:
 (i) an inhomogeneous polar distortion propagating into the
 bulk of the film, (ii) the film becoming partially or totally metallic 
 due to a non-vanishing free charge, and (iii) the local conduction
 band edge crossing the Fermi level.
 The above three effects are intimately linked, and should be considered
 as potential artifacts of the aforementioned band-gap problem.
 Whenever one of these ``alarm flags'' is raised in a calculation, the
 results should be examined with great caution.

 A route to overcoming this limitation involves correcting the LDA/GGA
 bandgap while preserving the excellent accuracy of these functionals in the
 prediction of ground-state properties.
 Traditional methods to increase the band gap of insulators, like
 the inclusion of a Hubbard $U$ term in the Hamiltonian,
 are not satisfactory in the case of a ferroelectric capacitor with
 B-cation driven ferroelectricity. 
%
%
 A more promising avenue has been recently opened
 by Bilc {\it et al.}~\cite{Bilc-08} and Wahl and coworkers,~\cite{Wahl-08}
 using the so-called ``hybrid'' functionals that combine Hartree-Fock
 exchange and DFT.
%
%
 In particular the B1-WC functional proposed in
 Ref. \onlinecite{Bilc-08} 
 has been shown to provide good structural, electronic and ferroelectric 
 properties as compared to experimental data for BaTiO$_{3}$ and PbTiO$_{3}$; 
 verifying the accuracy of B1-WC in interface studies will be an interesting
 subject for future research.
%
 Unfortunately, the price to pay for this accuracy is the substantially higher  
 computational cost of B1-WC compared to LDA/GGA.

%
%

 In addition to the purely technical issues, our work also opens interesting
 avenues regarding fundamental physical concepts. For example, ferroelectricity
 is usually understood within the modern theory of polarization, which is only
 applicable in the absence of conduction electrons (i.e. in pure insulators at
 zero electronic temperature).
 It is an important fundamental question, therefore, to assess whether our
 understanding of ferroelectrics in terms of bound charges, polarization and
 macroscopic electrical quantities still applies (and to what extent) in a
 regime where a sizable amount of space charge is present in the system.
 This issue is of crucial importance also for other systems, e.g. electrostatically
 doped perovskites, which bear many analogies to the physical mechanisms discussed 
 in this work.
 The first-principles-based modeling approach proposed in Ref.~\onlinecite{laosto}
 appears to be a promising route to further exploring this interesting topic.

\section{Acknowledgements}
 The authors are indebted to Ph. Ghosez and A. Filippetti for useful 
 discussions.
 This work was supported by the Spanish Ministery of Science and
 Innovation through the MICINN Grant FIS2009-12721-C04-02 (JJ); by the 
 Spanish Ministry of Education through the FPU fellowship AP2006-02958 (PAP);
 by DGI-Spain through Grants No. MAT2010-18113 and No. CSD2007-00041 (MS);
 by the European Union through the project EC-FP7, 
 Grant No. NMP3-SL-2009-228989 ``OxIDes'' (JJ and MS);
 and by the US National Science Foundation, award number DMR-0940420 (NAS).
 JJ, PAP and MS thankfully acknowledge the computer resources, 
 technical expertise and assistance provided by the 
 Red Espa\~nola de Supercomputaci\'on.
 Calculations were also performed at the ATC group
 of the University of Cantabria, and at CESGA.

\appendix

\section{Local polarization via Born effective charges}
\label{app:a}

 In this Appendix we discuss the approach, used in several parts in this
 manuscript and ubiquitously in the recent literature, of associating the
 local value of the ``effective'' polarization (i.e. the induced $P$
 with respect to the reference centrosymmetric configuration~\cite{ferro:2007}) 
 in capacitor heterostructures with an approximate formula, based on the 
 Born effective charges, $Z^*$.
 In particular, we provide formal justification for an improved formula, 
 still based on the $Z^*$, that we introduced in this work, and we already 
 mentioned in Sec.~\ref{sec:localpolborn}.

 Recall the definition of the approximate effective polarization in terms 
 of the Born effective charges in a bulk solid,
 
 \begin{equation}
    P^Z = \frac{e}{\Omega}
    \sum_{\alpha} Z^{\ast}_\alpha R_{\alpha z}.
    \label{eq:polbec}
 \end{equation}

 \noindent It is easy to verify that the layer-resolved 
 expression $P_j^Z$ of Eq. (\ref{eq:localpolbec3})
 reduces to $P^Z$ in the case of a periodic crystal,
 where $P_j^Z$ is a constant function of the
 layer index $j$.
 $P^Z$ does not reduce to the ``correct'' polarization $P(D)$
 at any value of $D$, as it
 does not take into account the additional polarization of the
 electronic cloud due to the
 internal field $\mathcal{E}(D)$
 (recall that the Born effective charges are defined
 under the condition of
 \emph{zero macroscopic electric field}.~\cite{Ghosez-98})
 \begin{table}
    \begin{ruledtabular}
       \begin{tabular}{cccc}
                                          & 
          $\epsilon_{\rm TOT}$            & 
          $\epsilon_\infty$               & 
          $\chi_{\rm TOT}/\chi_{\rm ION}$ \\
\hline
          BaTiO$_3$ & -48.87  &  6.48  & 0.90 \\
          PbTiO$_3$ & -96.54  &  8.33  & 0.93 \\
          KNbO$_3$  & -34.92  &  6.27  & 0.87 
      \end{tabular}
    \caption{ \label{tab:chi} Values of the susceptibilities $\chi$ and
              scaling factors $\chi_{\rm TOT}/\chi_{\rm ION}$ 
              for the ferroelectric materials considered in this work.}
    \end{ruledtabular}
 \end{table}

 Taking the Taylor expansion of the polarization as a function of $D$
 (we assume for simplicity that $D$, $P$ and $P^Z$ all vanish in the 
 reference centrosymmetric structure), we can write


 \begin{equation}
    P^Z(D)  = \frac{d P^Z}{d D} D + \ldots    
            = \frac{d P^Z}{d \mathcal{E}} 
               \frac{d \mathcal{E}}{d D} D + \ldots    
    \label{eq:taylorpjd}
 \end{equation}

 \noindent For small values of $D$, we can truncate the previous expansion at
 the linear order term.
 Now, by definition

 \begin{equation}
    \frac{d P^Z}{d \mathcal{E}} = \chi_{\rm ION},
    \label{eq:chiion}
 \end{equation}

 \noindent where $\chi_{\rm ION}$ is the lattice-mediated susceptibility, and
 
 \begin{equation}
    \frac{d \mathcal{E}}{d D} = (\epsilon_0 \epsilon_{\rm TOT})^{-1},
    \label{eq:epsilontot}
 \end{equation}

 \noindent where $\epsilon_{\rm TOT}$ is the total dielectric 
 constant of the insulator (relative to the vacuum permittivity $\epsilon_0$). 
 Substituting Eq. (\ref{eq:chiion}) and Eq.~(\ref{eq:epsilontot}) into 
 Eq.~(\ref{eq:taylorpjd})

 \begin{equation}
    P^Z(D) \sim D \frac{\chi_{\rm ION}}{\epsilon_0 \epsilon_{\rm TOT}}.
 \end{equation} 

 The same kind of arguments applied to the \emph{total} polarization 
 yield 

 \begin{equation}
    P(D) \sim D \frac{\chi_{\rm TOT}}{\epsilon_0 \epsilon_{\rm TOT}},
 \end{equation}

 \noindent where $\chi_{\rm TOT}$ is the sum of the lattice-mediated 
 susceptibility, $\chi_{\rm ION}$, and the purely electronic (frozen-ion)
 susceptibility, $\chi_{\infty}$.
 Note that $\chi_{\rm ION}$ is not bound to be positive. 
 In a ferroelectric material, for example, the centrosymmetric
 reference structure is unstable and therefore yields a
 negative $\chi_{\rm ION}$ (and hence $\epsilon_{\rm TOT}$),
 as discussed in Ref.~\onlinecite{fixedd}.
 The present derivation is general and encompasses those cases.

 \begin{figure}
    \begin{center}
       \includegraphics[width=3.2in,clip] {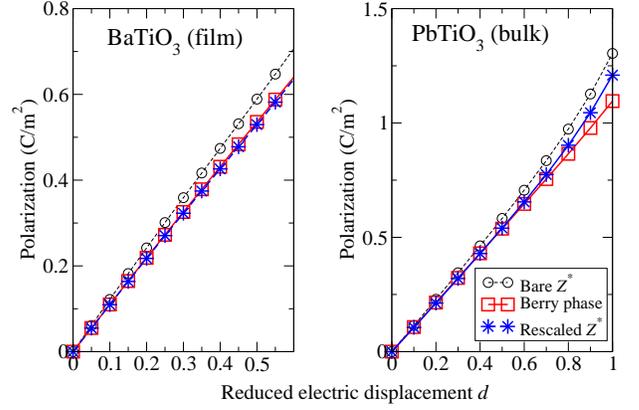}
       \caption{Polarization $P$ in a BaTiO$_3$ \emph{film} and
                 PbTiO$_3$ \emph{bulk} as a 
                function of the reduced electric 
                displacement field $d=DS$. 
                Data are taken from
                Ref.~\onlinecite{Stengel-09.2} (see Section III.C.1)
                and Ref.~\onlinecite{fixedd}.}
       \label{fig:zstar_bto}
    \end{center}
 \end{figure}

 From the above considerations it immediately follows that an estimate of
 the total polarization, which is \emph{exact} in the linear limit, can
 be given as 
 
 \begin{equation}
    P(D) \sim \frac{\chi_{\rm TOT}}{\chi_{\rm ION}} P^Z(D).
 \end{equation}
 This is essentially Eq. (\ref{eqzstar}). In practice, $\chi_{\rm ION}$ and 
 $\chi_{\infty}$ are calculated in the reference phase according to the
 standard definitions,~\cite{Gonze-97}

 \begin{equation}
    \chi_{\rm ION} = \epsilon_0 (\epsilon_{\rm TOT} - \epsilon_\infty) = 
    \frac{\epsilon_0 e^2}{M_0 \Omega} \sum_m \frac{(\tilde{Z}^*_m)^2}{\omega^2_m},
 \end{equation}

 \noindent where $M_0$ is a unit mass, $\tilde{Z}^*_m$ are the 
 normal mode charges
 and $\omega^2_m$ are the eigenvalues of the dynamical matrix, and

 \begin{equation}
    \chi_{\infty} = \epsilon_0 (\epsilon_{\infty}-1), \qquad  
    \epsilon_{\infty}^{-1} = \epsilon_0
    \frac{d\mathcal{E}}{dD} \Big|_{\rm fixed-ions}.
 \end{equation}
 The values of these physical constants that are relevant for the results
 presented in this manuscript are reported in Table~\ref{tab:chi}.


 We proceed in the following to test this approximation on two
 representative bulk ferroelectric materials, PbTiO$_3$ and BaTiO$_3$. 
 We take the relevant data
 (linear susceptibilities, Born charges and relaxed structures as a function of 
 $D$) from the calculations of Ref.~\onlinecite{fixedd} and
 Ref.~\onlinecite{Stengel-09.2}.
 Note that the BaTiO$_3$ calculation was performed at a fixed value of
 the in-plane lattice
 parameter (indicated as ``film'' in the figure) while in the
 PbTiO$_3$ calculation 
 both $a$ and $c$ parameters were relaxed for each value of $D$.
 The results are presented in Fig.~\ref{fig:zstar_bto}.
 In both cases, the ``bare'' value $P^Z$ is systematically overestimated
 compared to the Berry-phase polarization. 
 With the correction described above, i.e. by rescaling all values by the factor
 $\chi_{\rm TOT}/\chi_{\rm ION}$, the approximate value of $P$ accurately
 matches the Berry-phase one.
 The accuracy is surprisingly good in BaTiO$_3$, where the maximum deviation is
 of the order of 1\%. In PbTiO$_3$, for large values of $d$ the rescaled-$Z^*$ 
 value of $P$ presents significant deviations. Note that these deviations
 mostly concern values of $d$ that are larger than that of the ferroelectric 
 ground state ($d\sim 0.74$), and therefore are not of concern in this manuscript.
 We ascribe these deviations to the field-induced structural transition that
 was described in Ref.~\onlinecite{fixedd}. 

 In conclusion, this simple rescaling factor appears to be an effective way
 to obtain a relatively accurate value of the local $P$ in heterostructure
 calculations,
 based only on the local atomic positions and a few ingredients that can be
 easily computed in the bulk reference structure. From the results of our
 tests, we expect the agreement to be best in cases where the polarization is 
 small (closer to the linear limit where the approximation becomes exact). 
 Furthermore, cases where the ferroelectric polarization can be represented
 in terms
 of a single ``soft mode'' such as BaTiO$_3$ seem to work better than cases, like
 PbTiO$_3$, where significant mode mixing and non-trivial structural transitions
 occur at higher $D$ values.

\section{Convolutions and energy smearing of the 
local density of states}
\label{app:b}

\subsection{Convolutions}

Convolution is a mathematical operation on two functions $f$ and $g$, 
producing a third function that is typically viewed as a modified version 
of one of the original functions.
For the purpose of the present notes, it is useful to think
of $f$ as a data curve containing the relevant physical information, and
$g$ as a rapidly-decaying ``smoothing'' function that produces a local weighted 
average of $f$. 
We define the convolution of $f$ and $g$, $f * g$, as the following integral
transform,
\begin{equation}
(f*g) (x) = \int_{-\infty}^{+\infty} f(y) g(x-y) dy
\end{equation}

Convolutions have many properties, including commutativity and
associativity. Furthermore, the Dirac delta can be thought as the 
identity under the convoluton operation,
\begin{equation}
(f*\delta)(x) = f(x),
\end{equation}
and under certain assumptions an inverse operation can also be defined.
In other words, the set of invertible distributions forms an abelian
group under the convolution.

A particularly useful property holds in relationship to the Fourier 
transform,
\begin{equation}
\mathcal{F}(f*g) = k \cdot \mathcal{F}(f) \cdot \mathcal{F}(g)
\label{eq:fourier}
\end{equation}
where $\mathcal{F}(f)$ denotes the Fourier transform of $f$,
and $k$ is a constant that depends on the normalization convention
for the Fourier transform.
Thus, in reciprocal space the convolution becomes a simple product. This
naturally provides an efficient convolution algorithm: the workload
is reduced from $\mathcal{O}(N^2)$ to $\mathcal{O}[N \log(N)]$.

\subsection{Local density of states}

In this work we use [Eq.~(\ref{eq:discrete})] the following formula to 
compute the smeared local density of states (LDOS),
 \begin{equation}
    \tilde{\rho}({\bf r}, E) = \sum_{n{\bf k}} w_{\bf k} 
                       \left|  \psi_{n {\bf{k}}}({\bf r})  \right|^{2}
                       g(E - E_{n {\bf{k}}}).
    \label{eq:discretea}
 \end{equation}
We shall see that this is indeed a convolution. We first
get rid of the spatial cordinates. To this end, it is customary to 
integrate the LDOS in real space over a given volume $V$,
 \begin{equation}
    \rho_V(E) = \sum_{n{\bf k}} w_{\bf k} 
                       \rho_{n {\bf{k}}}(V)
                       g(E - E_{n {\bf{k}}}),
 \end{equation}
where
 \begin{equation}
    \rho_{n {\bf{k}}}(V) = \int_V d^3r \left|  \psi_{n {\bf{k}}}({\bf r})  \right|^{2}.
 \end{equation}
(Note that sometimes it might be more convenient to use a projected
density of states, rather than a local density of states. In such cases
it is sufficient to replace the real-space integral in the above equation
with an appropriate sum over angular momentum components. The following 
discussion remains unchanged.)
Now the LDOS is a function of a single energy variable. If we write
\begin{equation}
f_V(E) = \sum_{n{\bf k}} w_{\bf k} 
                       \rho_{n {\bf{k}}}(V)
                       \delta(E - E_{n {\bf{k}}}), 
\end{equation}
we can easily see that $\rho_V = f_V * g$. This leads to a simple reciprocal-space
expression. We first define an energy window, $[E_{\rm low},E_{\rm high}]$, that 
contains the entire eigenvalue spectrum $E_{n {\bf{k}}}$. We actually take a 
window which is slightly larger, where this ``slightly'' depends on the decay 
properties of $g$,
\begin{equation}
E_{\rm low} =  \min (E_{n {\bf{k}}}) - \epsilon, \qquad 
E_{\rm high} = \max (E_{n {\bf{k}}}) + \epsilon.
\end{equation}
The width of this window is $E_{\rm high}-E_{\rm low}=\Delta E$. We represent
$\rho_V(E)$ in reciprocal space as a discrete Fourier transform,
\begin{equation}
\rho_V(E) = \sum_\omega e^{i\omega E} \rho_V(\omega),
\label{eq:recip}
\end{equation}
where $\omega = 2\pi n/ \Delta E$ and $n$ is an integer. By using 
Eq.~(\ref{eq:fourier}) we have
\begin{equation}
\rho_V(\omega) = \Delta E \cdot f_V(\omega) \cdot g(\omega).
\label{eq:product}
\end{equation}
The Fourier transform of a Dirac delta centered in the origin is a
constant. Eq.~(\ref{eq:product}) then decomposes the local density of
states into a \emph{structure factor}, 
\begin{equation}
f_V(\omega) = \frac{1}{\Delta E} \sum_{n{\bf k}} w_{\bf k} 
                       \rho_{n {\bf{k}}}(V) e^{-i \omega E_{n{\bf k}}},
\end{equation}
and a \emph{form factor} $g(\omega)$. Obviously, this  
formulation is only convenient if the function $g$ has a fast decay in
both real and reciprocal space, so that the sum in Eq.~(\ref{eq:recip})
can be truncated. This is indeed the case for the most widely used 
smoothing functions $g$, as we shall see in the following.

\subsection{Gaussian vs. Fermi-Dirac smearing}

\begin{figure}
 \begin{center}
 \includegraphics[width=3.0in,clip]{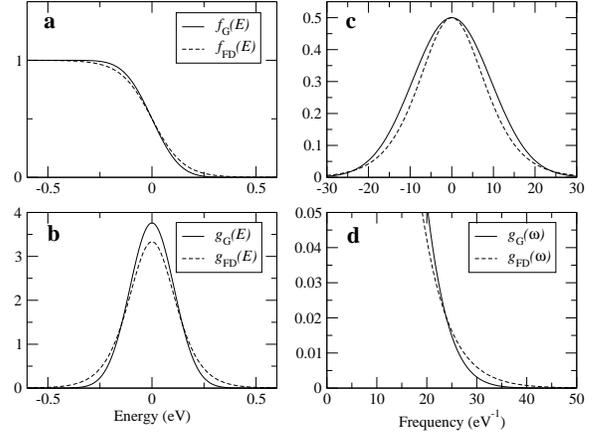}
 \end{center}
 \caption{(a) Gaussian ($\sigma = 0.15$ eV) and Fermi-Dirac 
   ($\sigma = 0.075$ eV) occupation functions. (b) Kernel of the 
   occupation functions as defined in the text. (c-d) Fourier 
   transform of the smearing kernels $g$, assuming an energy 
   window of $[-1,1]$. \label{fig:smear}}
\end{figure}

 The Gaussian smearing (G) and the Fermi-Dirac (FD) smearing are by far the 
 most popular choices for the occupation function in first-principles
 calculations of metallic systems. If we define the occupation function 
 $f$ as the integral of a ``kernel'' function $g$,
 \begin{equation}
   f(E) = 1 - \int_{-\infty}^E g(x) dx,
  \label{eq:ff}
 \end{equation}
 one can verify that the Gaussian or Fermi-Dirac occupation are, 
 respectively, reproduced by the following choices of $g$,
 \begin{eqnarray}
 g_{\rm G}(x) & = & \frac{1}{\sqrt{\pi} \sigma} e^{-x^2/\sigma^2},
       \label{eq:gga}\\
 g_{\rm FD}(x) & = & \frac{\sigma^{-1}}{2+e^{x/\sigma} + e^{-x/\sigma}},
       \label{eq:gfda}
  \end{eqnarray}
where $\sigma$ is the smearing energy [these correspond to 
Eq.~(\ref{eq:gg}) and Eq.~(\ref{eq:gfd})].
It is easy to see that, by combining Eq.~(\ref{eq:gga}) or Eq.~(\ref{eq:gfda}) 
with Eq.~(\ref{eq:ff}) one obtains the standard definitions of the occupation
function (we assume that the complementary error function, erfc, values
2 at $-\infty$),
 \begin{eqnarray}
 f_{\rm G}(x) & = \frac{1}{2} \, {\rm erfc} \, (x/\sigma),
       \label{eq:fg}\\
 f_{\rm FD}(x) & = \frac{1}{e^{x/\sigma} + 1}.
       \label{eq:ffd}
 \end{eqnarray}
 It is useful to spell out the explicit 
 formulas for the Fourier transforms of both smearing functions,
 \begin{eqnarray}
 g_{\rm G}(\omega) & = & \frac{e^{-\omega^2 \sigma^2 / 4}}{\Delta E} ,
       \label{eq:ggom}\\
 g_{\rm FD}(\omega) & = & \frac{\pi \omega \sigma}{ \Delta E \, \sinh (\pi \omega \sigma)}.
       \label{eq:gfdom}
 \end{eqnarray} 
 Note that the above formulas are normalized according to the 
 conventions on the Fourier transforms that we used in the previous
 section.
 The functions $f$ and $g$ defined above are shown in Fig.~\ref{fig:smear}.
 Note that a different choice of $\sigma$ was used in the Fermi-Dirac
 and in the Gaussian case. A FD distribution is roughly equivalent to
 a G distribution with a $\sigma$ value that is twice as large.
 
 In the main text and here we have assumed that it is a good idea to
 use the same $g$ kernel in the calculation and in the construction of
 the LDOS.
 We shall substantiate this point in the following Section.

\subsection{On the optimal choice of $g$}

\begin{figure}
 \begin{center}
 \includegraphics[width=3.0in,clip]{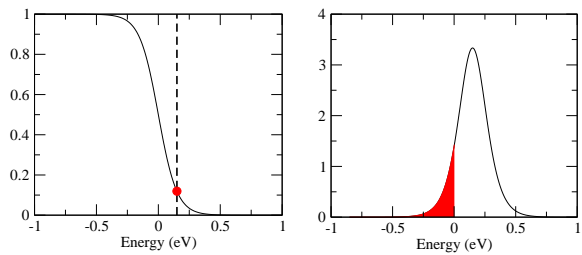}
 \end{center}
 \caption{Left: Fermi-Dirac occupation function, identical to that of
  Fig.~\ref{fig:smear}(a) (solid curve); hypothetical orbital located at 
  an energy of 0.15 eV above the Fermi level (dashed line); the thermal 
  occupation of this state yields a total charge of 0.119 electrons (red dot). Right: 
  density of states corresponding to the single isolated orbital at an energy of
  0.15 eV above the Fermi level, smeared by using the $g_{\rm FD}$ kernel
  of Fig.~\ref{fig:smear}(b); the integral of the DOS up to the Fermi
  level (shaded area) yields the exact same charge of 0.119 electrons. \label{fig:occup}}
\end{figure}   

In many cases, the specific choice of the $g$ function to be used in
Eq.~(\ref{eq:discretea}) is largely arbitrary. Typically, the goal is
to filter out the unphysical wiggles due to the discretization of the $k$-mesh,
but at the same time to preserve the main physical features, without
blurring them out completely. This calls for a smearing function that is neither
too sharp nor too broad. Since a ``slightly too broad'' or a ``slightly too
sharp'' smearing function usually does not influence the physical conclusions,
in many cases one has the freedom of choosing whatever yields the clearest 
visual aid to support the discussion.

There are cases, however, where this choice is not just a matter of aesthetics,
and using the ``wrong'' $g$ function can qualitatively and quantitatively 
influence the interpretation of the results. 
More specifically, the issue concerns cases where the analysis of the
LDOS (or DOS or PDOS) is used to detect and quantify the population of
orbitals that lie close in energy to the Fermi level.
As we focus on charge spill-out phenomena that concern the conduction
band of a dielectric/ferroelectric film in contact with a metallic 
electrode, this is a central point of our work.
The problem is most easily appreciated by looking at the left panel of 
Fig.~\ref{fig:occup}. There is a single orbital lying at an energy of
0.15 eV above the Fermi level. 
As this orbital lies \emph{above} the Fermi level, one might be tempted to
think that the orbital is empty, and that charge spill-out does not occur at
all.
However, calculations in metallic systems are routinely performed by 
using an occupation function that is artificially broadened, in order to
improve convergence of the ground-state properties; in Fig.~\ref{fig:occup}
we assume a Fermi-Dirac occupation with a fictitious electronic temperature of
0.075 eV.
It is easy to see that with such an occupation function, the orbital lying at
0.15 eV won't be empty, but will be ``thermally'' populated by tail of the
Fermi-Dirac distribution. The final result is a charge transfer of 0.119 
electrons into this orbital.

Now, is there a ``right'' way to construct the DOS curve, such that the
above-mentioned charge transfer could be qualitatively and quantitatively 
inferred from the DOS, without knowing any further detail of the calculation?
The answer is yes, and consists in constructing the DOS by using as broadening
$g$ function which is consistent with the occupation function used by the
code. 
In this case, this is $g_{\rm FD}$, with a $\sigma$ identical to that used
to calculate the electronic ground state.
To demonstrate this point, we plot in the right panel of Fig.~\ref{fig:occup}
the DOS of this isolated orbital at 0.15 eV, appropriately convoluted with
$g_{\rm FD}$.
Eq.~(\ref{eq:ff}) guarantees that, by doing this, one recovers the very intuitive result
that the total amount of electron charge, $Q$, present in the volume $V$ (over which 
the LDOS was integrated) \emph{exactly} corresponds to the integral of the DOS
up to the Fermi level,
\begin{equation}
Q = \int_{-\infty}^{E_{\rm F}} \rho_V(E) dE.
\end{equation}
Then, a simple look at the DOS curve is sufficient to ascertain whether 
a significant transfer of charge has occurred into a specific group of
bands.
As this rigorous sum rule can be very practical in the analysis of the
results, we encourage a systematic use of the ``internally consistent'' LDOS
construction described above.

\bibliography{max-feb28}

\begin{thebibliography}{77}
\bibitem[{k-p()}]{k-point-sampling}
\bibinfo{note}{In practical simulations, the origin of the $k$-point grid may
  be displaced from $k$ = 0 in order to decrease the number of inequivalent
  $k$-points. \cite{Monkhorst-76, Moreno-92} This shift usually prevents the
  appearance of high-symmetry points from the list of $k$-points used during
  the self-consistent procedure or in the calculations of the density of
  states.}
\bibitem[{Ban()}]{Band-gap-BTO-PTO}
\bibinfo{note}{The band gap of both BaTiO$_{3}$ and PbTiO$_{3}$ is indirect,
  with the top of the valence band located at $R$ in BaTiO$_{3}$ and at $X$ in
  PbTiO$_{3}$, and the bottom of the conduction band at $\Gamma$ in both
  materials.}
\end{thebibliography}
\end{document}